\documentclass[aps,prl,twocolumn,superscriptaddress]{revtex4-1}

\usepackage{graphicx}

\usepackage{algorithm}
\usepackage{color}
\usepackage{algpseudocode}
\usepackage{amsmath,amsfonts,amsthm} 
\usepackage{verbatim} 

\usepackage{multirow}
\usepackage{epstopdf}
\usepackage{natbib}

\definecolor{darkblue}{rgb}{1.0,0.0,0.5}

\begin{document}


\title{Extrapolating quantum observables with machine learning: Inferring multiple phase transitions from properties of a single phase}


\author{Rodrigo A. Vargas-Hern\'andez$^{1}$, John Sous$^{1,2,3},~$Mona Berciu$^{2,3}$, and Roman V. Krems}
\affiliation{Department of Chemistry, University of British Columbia, Vancouver, British Columbia, Canada, V6T 1Z1\\
$^2$Department of Physics and Astronomy, University of British Columbia, Vancouver, British Columbia, Canada, V6T 1Z1\\
$^3$Stewart Blusson Quantum Matter Institute, University of British Columbia, Vancouver, British Columbia, Canada, V6T 1Z4\\
}


\date{\today}

\begin{abstract}
We present a machine-learning method for predicting sharp transitions in a 
Hamiltonian phase diagram by extrapolating the properties of quantum systems. 
The method is based on Gaussian Process regression with a combination of 
kernels chosen through an iterative procedure maximizing the predicting power of the kernels.
The method is capable of extrapolating across the transition lines. 
The calculations within a given phase can be used 
to predict not only the closest sharp transition, but also a transition removed from the available data by a separate phase.
This makes the present method particularly valuable for searching phase transitions in the parts of the parameter space that cannot be probed experimentally or theoretically. 

\end{abstract}

\pacs{}

\maketitle


It is very common in quantum physics to encounter a problem with the Hamiltonian
\begin{eqnarray}
 H =  H_0 + \alpha H_1 + \beta  H_2
\label{model}
\end{eqnarray}
whose eigenspectrum can be readily computed/measured in certain limits of $\alpha$ and $\beta$, e.g. at $\alpha=0$ or at $\alpha \gg \beta$, but not at arbitrary values of $\alpha$ and $\beta$. 
For such problems, it is necessary to interpolate the properties of the quantum system between the known limits or extrapolate from a known limit. Both the interpolation and extrapolation become exceedingly complex if the system properties undergo sharp transitions at some values of $\alpha$ and/or $\beta$.  Such sharp transitions separate the phases of the Hamiltonian (\ref{model}).
Because the {wave functions} of the quantum system are drastically different in the different phases \cite{sachdev}, an extrapolation of quantum properties across phase transition lines is generally considered unfeasible. 

Here, we challenge this premise. {We note that, while certain properties of quantum systems undergo a sharp change at a phase transition, other properties evolve smoothly through the transition. 
Using the example of three different lattice models, we show that the evolution of such properties within a given phase contains information about the transitions and the same properties beyond the transitions. We present a machine-learning method that can be trained by {\it the evolution} of such properties in a given phase to predict the sharp transitions and the properties of the quantum system in other phases by extrapolation.}
 The importance of this result is clear. Characterizing quantum phase transitions embodied in model Hamiltonians is one of the foremost goals of quantum condensed-matter physics. Our work illustrates the possibility of predicting transitions at Hamiltonian parameters, where obtaining the solutions of the Schr\"{o}dinger equation may not be feasible.


The application of machine learning (ML) tools for solving problems in condensed-matter physics has recently become popular 
{\cite{WangPRB2016,CarasquillaNatPhys2017,  NieuwenburgNatPhys2017, BroeckerarXiv2017, WetzelScherzerarXiv2017,Wetzel2017,LiuarXiv2017,ChangPRX2017, BroeckerSciRep2017, SchindlerPRB2017, OhtsukiJPSJapan2016, ArsenaultPRB2014, ArsenaultarXiv2015, BeacharXiv2017, RafaelarXiv2017, YoshiokaarXiv2017, VenderleyarXiv2017, RBM_Troyer, SchmittarXiv2017, CaiPRB, HuangarXiv, DengPRB, NomuraarXiv2017, DengPRX, GaoNatComm, TorlaiarXiv2017, GPdeepNN_2015, DanielyNIPS2016, GPdeepNN_2017,MLcrystal1, MLcrystal2, MLcrystal3, MLcrystal4}}.
 In all of these applications, ML is used as an efficient method to solve one of three problems: interpolation, classification or clustering.  
Interpolation amounts to fitting multi-dimensional functions or functionals, whereas classification and clustering are used to separate physical data by properties. 
For example, ML can be used to identify quantum phases of lattice spin Hamiltonians \cite{WangPRB2016,CarasquillaNatPhys2017,DengPRB,ChangPRX2017,RBM_Troyer,NomuraarXiv2017,RafaelarXiv2017}.  
However, in order to identify a quantum phase transition by interpolation and/or classification, the aforementioned ML models must be trained (fed on input) by the data describing both phases on both sides of the transition. The distinct feature of the present work is a ML method {that requires information from only one phase and {\it extrapolates} the properties of lattice models to and across the transitions.} 
{To illustrate the method, we consider {four different problems: lattice polaron models with zero, one and two sharp transitions}, and the mean-field Heisenberg model with a critical temperature. In all cases, we show that the phase transitions ({or lack thereof}) can be accurately identified. 
}



We first consider a generalized lattice polaron model describing an electron in a one-dimensional lattice with $N \rightarrow \infty$ sites coupled to a phonon field: 
 \begin{eqnarray}
{\cal H} =   \sum_k  \epsilon_k c^\dagger_k c_k +  \sum_q \omega_q  b^\dagger_q b_q + V_{\rm{e-ph}},
\label{polaron}
\end{eqnarray}
where $c_k$ and $b_q$ are the annihilation operators for the electron with momentum $k$ and phonons with momentum $q$,  $\epsilon_k = 2 t \cos(k)$ and $\omega_q = \omega = {\rm const}$ are the electron and phonon dispersions, and $V_{\rm{e-ph}}$ is the electron-phonon coupling. We choose $V_{\rm{e-ph}}$ to be a combination of two qualitatively different terms $V_{\rm{e-ph}} = \alpha H_1 + \beta H_2$, where 
\begin{eqnarray}
H_1 = \sum_{k,q} \frac{2i}{\sqrt{N}}\left [ \sin(k + q) - \sin(k) \right]c^\dagger_{k+q}c_k \left ( b^\dagger_{-q}  + b_q\right ) \quad
\end{eqnarray}
describes the Su-Schrieffer-Heeger (SSH) \cite{ssh} electron-phonon coupling, and  
\begin{eqnarray}
H_2 = \sum_{k,q} \frac{2i}{\sqrt{N}} \sin(q)c^\dagger_{k+q}c_k \left ( b^\dagger_{-q}  + b_q\right )
\end{eqnarray}
is the breathing-mode model \cite{breathing-mode}. The ground state band of the model (\ref{polaron}) represents polarons known to exhibit two sharp transitions as the ratio $\alpha/\beta$ increases from zero to large values \cite{HerreraPRL}. At $\alpha = 0$, the model (\ref{polaron}) describes breathing-mode polarons, which have no sharp transitions \cite{no-transition}. At $\beta = 0$, the model (\ref{polaron}) describes  SSH polarons, which exhibit one sharp transition in the polaron phase diagram \cite{ssh}. At these transitions, the ground state momentum and the effective mass of the polaron change abruptly.

{\it Method.} We use Gaussian Process (GP) regression as the prediction method \cite{gpbook}, {described in detail in the Supplemental Material \cite{Supp-0}.} 
The goal of the prediction is to infer an unknown function $f(\cdot)$ given $n$ inputs $\mathbf{x}_i$ and outputs $y_i$. 
The assumption is that $y_i = f(\mathbf{x}_i)$.  The function $f$ is generally multidimensional so $\mathbf{x}_i$ is a vector. 

GPs do not infer a single function $f(\cdot)$, but rather a distribution over functions, $p(f|{\bf X},\mathbf{y})$, 
where ${\bf X}$ is a vector of all known $\mathbf{x}_i$ and $\mathbf{y}$ is a vector of the corresponding values $y_i$. 
 This distribution is assumed to be normal. 
The joint Gaussian distribution of random variables $f(\mathbf{x}_i)$ is characterized by a mean $\mu(\mathbf{x})$ and a covariance matrix $K(\cdot,\cdot)$. The matrix elements of the covariance $K_{i,j}$ are specified by a kernel function $k(\mathbf{x}_i,\mathbf{x}_j)$ that quantifies the similarity relation between the properties of the system at two points $\mathbf{x}_i$ and $\mathbf{x}_j$ in the multi-dimensional space.

Prediction at $\bf x_*$ is done  by computing the conditional distribution of $f(\mathbf{x}_*)$ given $\mathbf{y}$ and ${\bf X}$. 
The mean of the conditional distribution is \cite{gpbook}
\begin{eqnarray}
\mathbf{\mu}(\mathbf{x}_*)  &=& \sum_i^{n} d(\mathbf{x}_*, \mathbf{x}_i)y_i  = \sum_i^{n} \alpha_i k(\mathbf{x}_*,\mathbf{x}_i)
\label{eqn:GP_pred}
\label{eqn:GP_std}
\end{eqnarray}
where ${\bf \alpha} = K^{-1}\mathbf{y}$ and $\mathbf{d} = K(\mathbf{x}_*,{\bf X})^\top K({\bf X},{\bf X})^{-1}$.
The predicted mean $\mathbf{\mu}(\mathbf{x}_*) $ can be viewed as a linear combination of the training data $y_i$ or as a linear combination of the kernels {connecting all} training points $\mathbf{x}_i$ and the point $\mathbf{x}_*$, where the prediction is made.  In order to train a GP model, one must choose an analytical representation for the kernel function. 

{To solve the {\it interpolation} problem, one typically uses a simple form for the kernel.}
In the limit of large $n$, any simple kernel function 
 produces accurate {interpolation} results \cite{gpbook}.
 For example, $k$ can be approximated by any of the following functions:  
\begin{eqnarray}
k_{\rm LIN}(\mathbf{x}_i, \mathbf{x}_j)  =  \mathbf{x}_i^\top \mathbf{x}_j
\label{eqn:k_LIN}
\end{eqnarray}
\begin{eqnarray}
k_{\rm RBF}(\mathbf{x}_i, \mathbf{x}_j) = \exp \left(-\frac{1}{2}r^2(\mathbf{x}_i, \mathbf{x}_j)\right)
\label{eqn:k_RBF}
\end{eqnarray}
\begin{eqnarray}
k_{\rm MAT}(\mathbf{x}_i, \mathbf{x}_j)  = \left( 1 + \sqrt{5}r^2(\mathbf{x}_i,\mathbf{x}_j) +  \frac{5}{3}r^2(\mathbf{x}_i, \mathbf{x}_j)\right )
\nonumber
\\
\times \exp\left ( -\sqrt{5}r^2(\mathbf{x}_i, \mathbf{x}_j)\right )~~~~
\label{eqn:k_MAT}
\end{eqnarray}
\begin{eqnarray}
k_{\rm RQ}(\mathbf{x}_i, \mathbf{x}_j)  = \left ( 1 + \frac{|\mathbf{x}_i- \mathbf{x}_j|^2}{2\alpha\ell^2} \right )^{-\alpha}
\label{eqn:k_RQ}
\end{eqnarray}
where $r^2(\mathbf{x}_i, \mathbf{x}_j) = (\mathbf{x}_i- \mathbf{x}_j)^\top \times {M} \times (\mathbf{x}_i-\mathbf{x}_j)$ and ${M}$ is a diagonal matrix with different length-scales $\ell_d$ for each dimension of $\mathbf{x}_i$. The length-scale parameters $\ell_d$, $\ell$ and $\alpha$ are the free parameters. We describe them collectively by $\mathbf{\theta}$. 
A GP is trained by finding the estimate of $\theta$ (denoted by $\hat \theta$) that maximizes the logarithm of the \emph{marginal likelihood} function:
\begin{eqnarray}
\log p(\mathbf{y} | \mathbf{X}, \mathbf{\theta}, {\cal M}_i) = -\frac{1}{2}\mathbf{y}^\top K^{-1}\mathbf{y} - \frac{1}{2}\log |K| - \frac{n}{2} \log 2\pi \nonumber \\
\end{eqnarray} 

{For the {\it extrapolation} problem, the prediction produced by} Eq. (\ref{eqn:GP_pred}) is clearly sensitive to the particular choice of the kernel function. 
{While different interpolation problems can be solved with the same mathematical form of the kernel function, different extrapolation problems generally require different kernels.  
The key for successful extrapolation is thus to find the appropriate kernel function. Because we aim to solve a variety of different problems with varying underlying physics, the procedure for constructing the kernel must be fully automated and independent of the particular problem under consideration.}


\begin{figure}[h!]
	\includegraphics[width=\columnwidth]{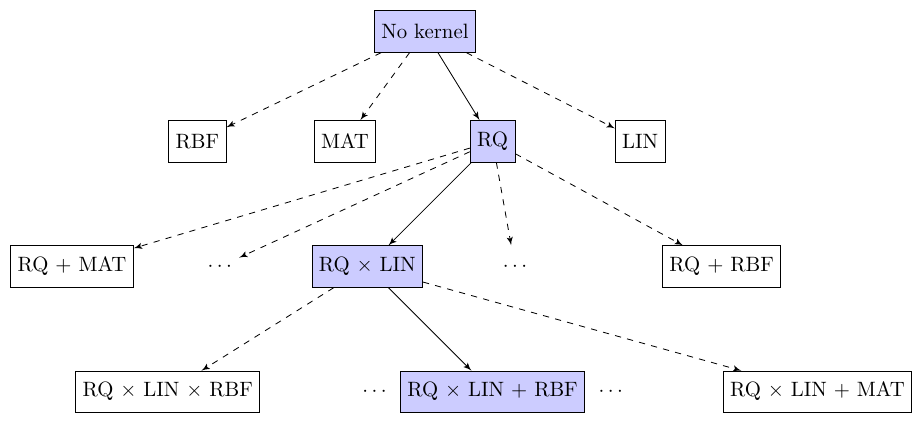}
	\caption{Schematic diagram of the kernel construction method employed to develop a Gaussian Process model with extrapolation power. At each iteration,
	the kernel with the highest Bayesian information criterion (\ref{BIC-eq}) is selected. {The labels in the boxes correspond to the kernel functions defined in (\ref{eqn:k_LIN})-(\ref{eqn:k_RQ})}.
	}
	\label{phases}
\end{figure}

\begin{figure*}[t]
	\includegraphics[width=\columnwidth]{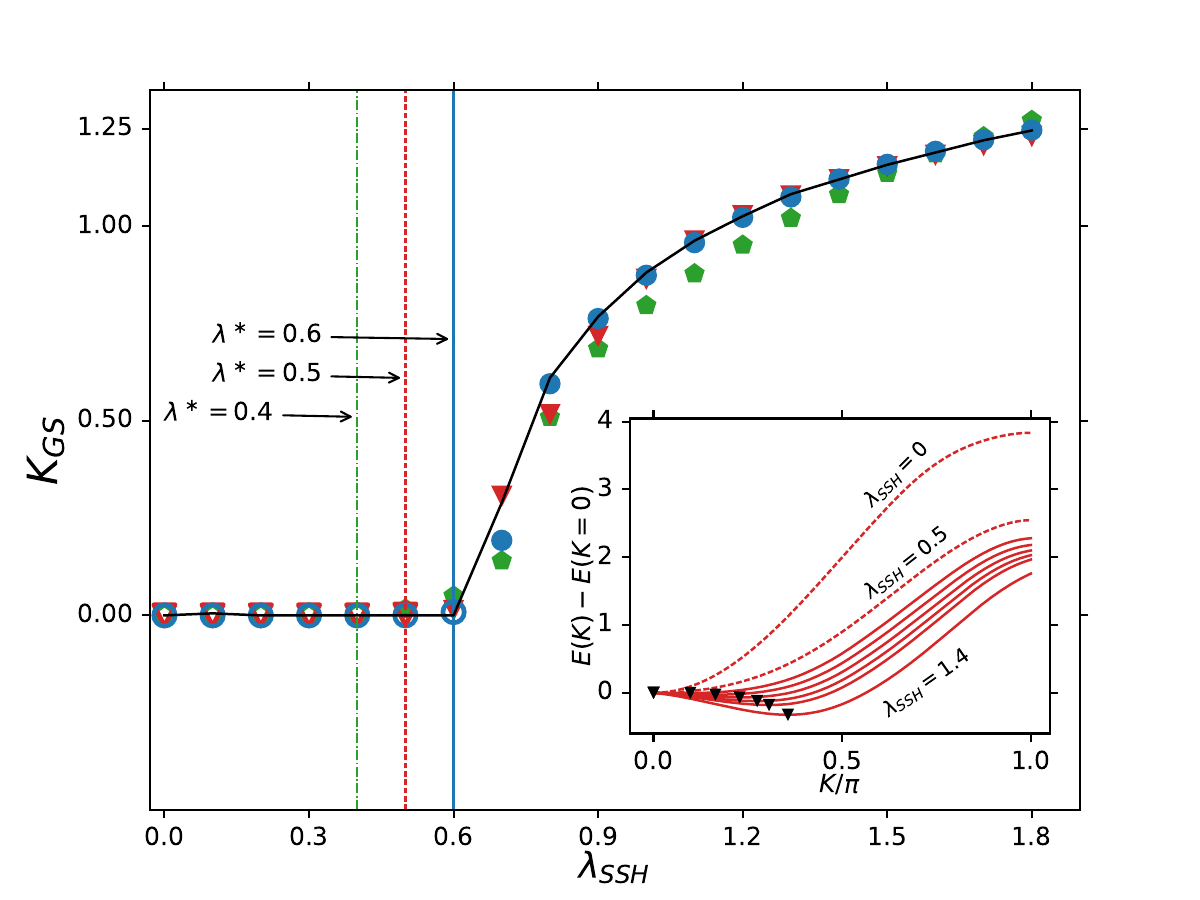}
	\includegraphics[width=\columnwidth]{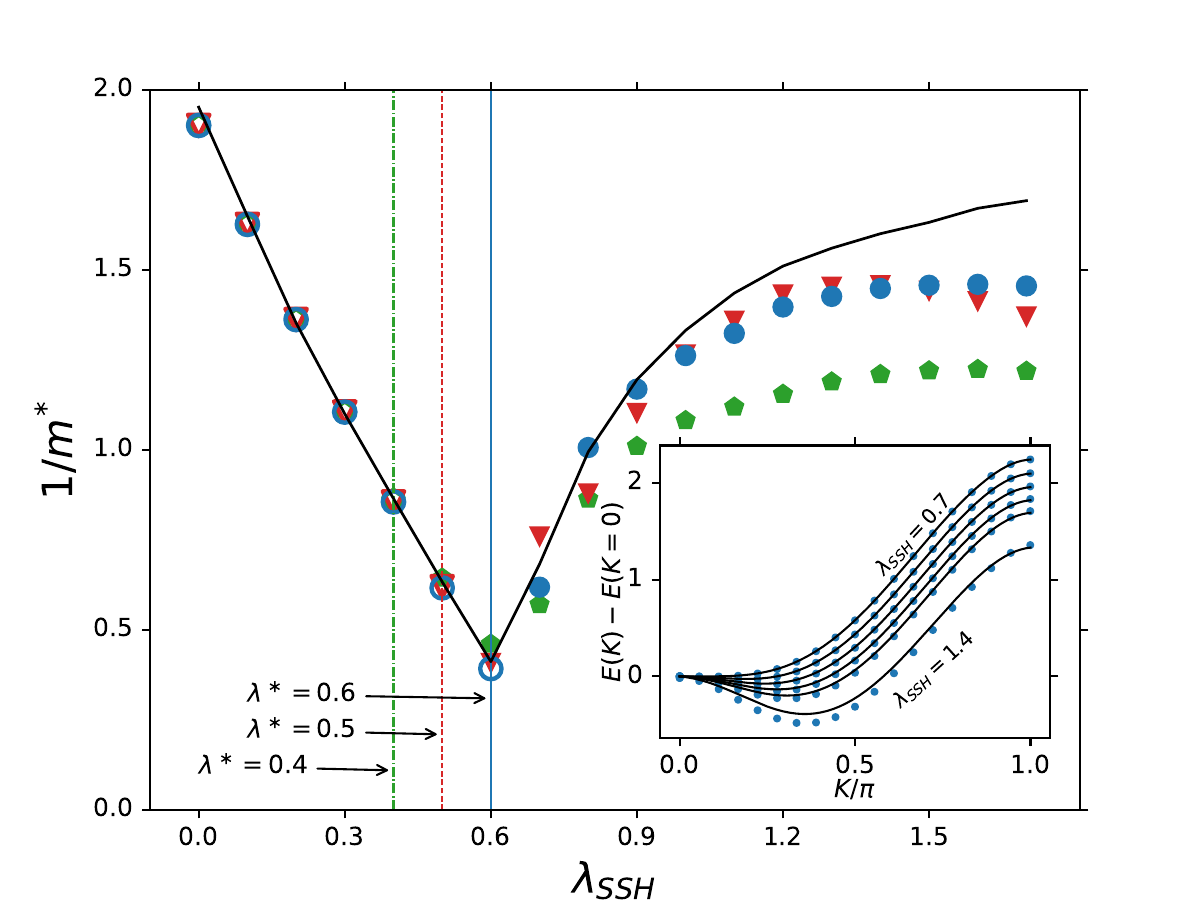}
	\caption{
	{Extrapolation of the polaron ground state momentum $K_{GS}$ (left) and effective mass $m^\ast$ (right) across the sharp transition at $\lambda_{SSH} = 2\alpha^2/t \hbar\omega \approx 0.6$ 
 for the model (\ref{polaron}) with $\beta = 0$.  The black solid curves are the accurate quantum calculations.  The symbols are the predictions of the GP models trained by the full polaron dispersions $E(K)$ at values of $\lambda_{SSH} \leq \lambda^\ast$, where $\lambda^\ast$ is shown by the vertical lines  (solid for circles, dashed for triangles and dot-dashed for pentagons).  The GP models are used for interpolation (open symbols) and extrapolation (full symbols). 
 The algorithm of Figure 1 yields the kernel  $k_{RQ}\times k_{LIN} + k_{RBF}$ for the GP models represented by the triangles and pentagons, and $k_{RQ}\times k_{LIN} \times k_{MAT}$ for the circles.
 Left inset: the polaron dispersions used as input (dashed curves) and predicted by the GP model (solid curves) with $\lambda^\ast = 0.5$ with the triangles showing the position of the dispersion minimum.  Right inset: the polaron dispersions predicted by the GP model trained with $\lambda^\ast = 0.6$ (solid curves) in comparison with the quantum calculations (symbols). 
%
	} 
}
	\label{phases}
\end{figure*}

Here, 
{we follow Refs. \cite{kernel_comb,gp-ss}}, to build a prediction method based on a {{\it combination} of products of different kernels (\ref{eqn:k_LIN})-(\ref{eqn:k_RQ})}. 
To select the best combination, we use the Bayesian information criterion (BIC) \cite{bic},
\begin{eqnarray}
\text{BIC}({\cal M}_i) = \log p(\mathbf{y} | \mathbf{X}, \hat{\mathbf{\theta}}, {\cal M}_i) -\frac{1}{2}|{\cal M}_i|\log n
\label{BIC-eq}
\end{eqnarray}
where $|{\cal M}_i|$ is the number of kernel parameters of kernel ${\cal M}_i$. 
 Here, $p(\mathbf{y} | {\bf X}, \hat{\mathbf{\theta}}, {\cal M}_i)$ is the marginal likelihood for an optimized kernel $\hat \theta$. 
It is impossible to train and try models with all possible combinations of kernels. We use an iterative procedure schematically depicted in Figure 1. We begin by training a GP model with  {\it each} of the kernels (\ref{eqn:k_LIN})-(\ref{eqn:k_RQ}). 
 These kernels have one (LIN), $d$ (RBF and MAT) and 2 (RQ) free parameters \cite{SM-kernelcombinations}. 
The algorithm then selects the kernel  -- denoted $k_0$ --  that leads to the model with the highest BIC and  
 combines $k_0(\cdot,\cdot)$ with each of the original kernels $k_i$ defined by Eqs. (6)-(9). 
The kernels are combined as products $k_0(\cdot,\cdot)\;\times \;k_i(\cdot,\cdot)$ and additions $k_0(\cdot,\cdot)\;+ \;k_i(\cdot,\cdot)$.  
Each kernel in the combination is scaled by a constant factor, which introduces another free parameter for the product or two parameters for the sum.
For each of the possible combinations, a new GP model is constructed and a BIC is computed. The kernel yielding the highest BIC is then used as a new base kernel $k_0$ and the procedure is iterated.  
This fully automated algorithm is applied here to four different problems, yielding physical extrapolation results, thus showing that Eq. (\ref{BIC-eq}) can be used 
as a criterion for building prediction models capable of physical extrapolation.

 


\begin{figure}[h!]
	\includegraphics[width=\columnwidth]{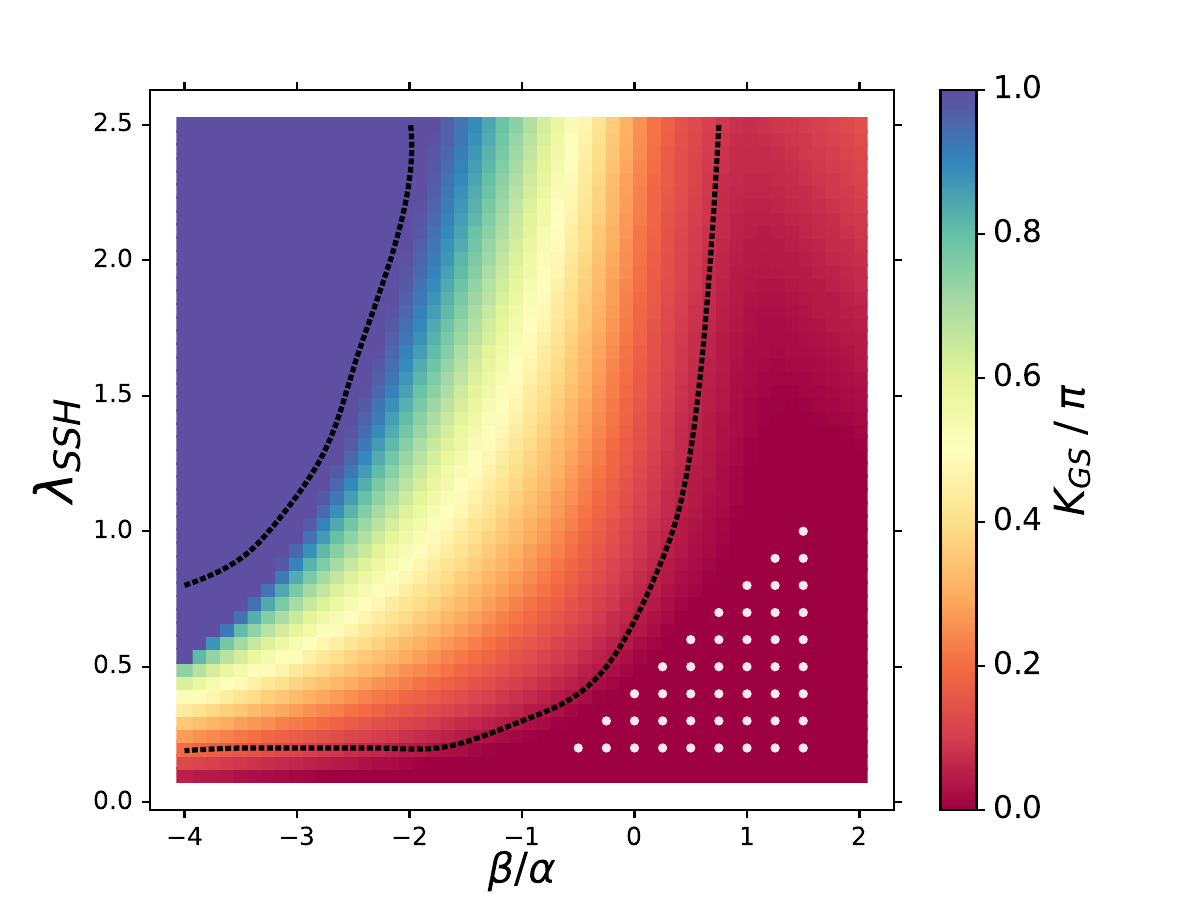}
	\includegraphics[width=\columnwidth]{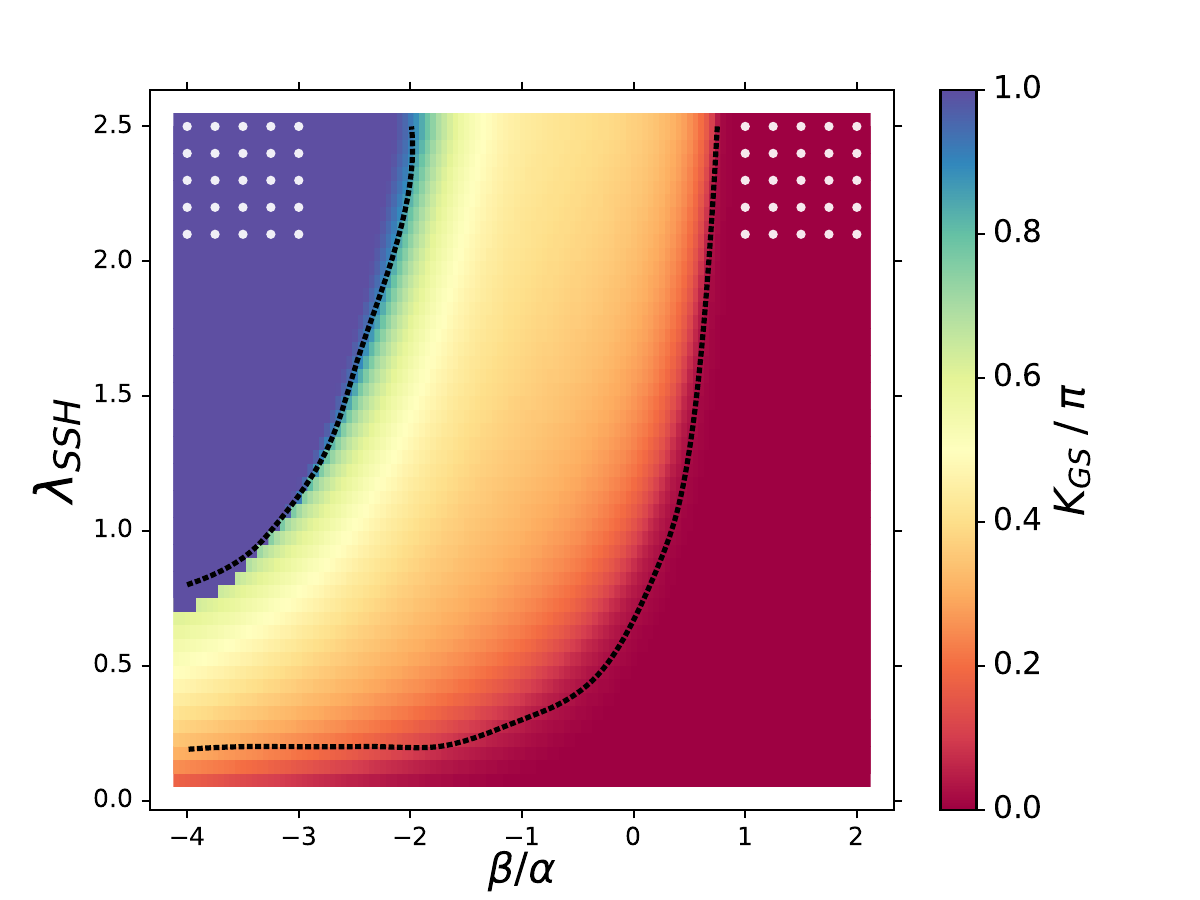}
	\caption{
	The polaron ground state momentum $K_{GS}$  for the mixed model (\ref{polaron}) as a function of $\beta/\alpha$ for $\lambda_{SSH} = 2\alpha^2/t \hbar\omega$. 
	The color map is the prediction of the GP models. 
	The curves are the quantum calculations from Ref. \cite{HerreraPRL}.  The models are trained by the polaron dispersions at the parameter values indicated by the white dots. No other information is used.
	{{The optimized kernel combination is $(k_{MAT} + k_{RBF})\times k_{LIN}$ (upper panel) and $(k_{MAT}\times k_{LIN} + k_{RBF})\times k_{LIN}$ (lower panel).}}}
	\label{phases}
\end{figure}

{\it Results.} All GP models are trained by the dispersions $E(K)$, where $E$ is the polaron energy and $K$ is the polaron momentum. 
These dispersions are calculated for {{infinite lattices using the Momentum Average (MA) approach}} from previous work \cite{MA,HerreraPRL,SousScRep,SousPRA,SousArxiv,SuppMA}. 
{The models are trained to predict} the polaron energy {as a function of $K$, and the Hamiltonian parameters $\alpha$, $\beta$, and $\omega$}. The vectors ${\bf x}_i$ are thus ${\bf x}_i \Rightarrow \{ K, \omega, \alpha, \beta \}$, {while $f(\cdot)$ is the polaron energy}. 
Once the models are trained,  we numerically compute the ground state momentum $K_{GS}$ and the polaron effective mass from the predicted dispersions \cite{Suppker}. {{Note that we always train all models by the polaron dispersions in one phase and the models have no {\it a priori} information about the existence of another phase(s). The transition is encoded in the evolution of the polaron band as a function of $\bf x$.}}  
All results are in units of $t$.

Figure 2 shows the predictions for the pure SSH polaron model ($\beta = 0$, one sharp transition in the polaron phase diagram).  
The vertical lines show where the training points end and the extrapolation begins. 
As can be seen, the GP models predict accurately the location of the transition and can be used for quantitative extrapolation in a wide range of the Hamiltonian parameters to strong electron-phonon coupling. {{All models, including the ones trained by quantum calculations far removed from the transition point, predict
accurately the location of the transition. As the coupling to phonons increases, the polaron develops a phonon-mediated next-nearest-neighbor hopping term: $E(K) = -2t {\rm{cos}} (K) + 2 t_2 (\lambda_{SSH}) {\rm{cos}} (2K)$, where $t_2(\lambda_{SSH})$ is a function of $\lambda_{SSH}$ \cite{ssh}. The transition occurs when the second term dominates. Figure 2 shows that the GP models trained using the algorithm of Figure 1 extrapolate accurately this evolution of the polaron energy. }}




 The power of this method is better illustrated with the example of the mixed breathing-mode -- SSH model ($\alpha \neq 0$, $\beta \neq 0$) with three phases \cite{HerreraPRL}. 
 The dots in Figure 3 represent the points of the phase diagram used for training the GP model with the optimized kernels. 
Remarkably, the model trained by the polaron dispersions all entirely in one phase predicts {\it both} transitions. The location of the first transition is predicted quantitatively. The second transition is predicted qualitatively. If the model is trained by the polaron properties in two side phases  and the prediction is made by extrapolation {to low values of $\lambda_{SSH}$} (lower panel of Figure 3), both transition lines are predicted quantitatively. 

\begin{figure}[h!]
	\includegraphics[width=\columnwidth]{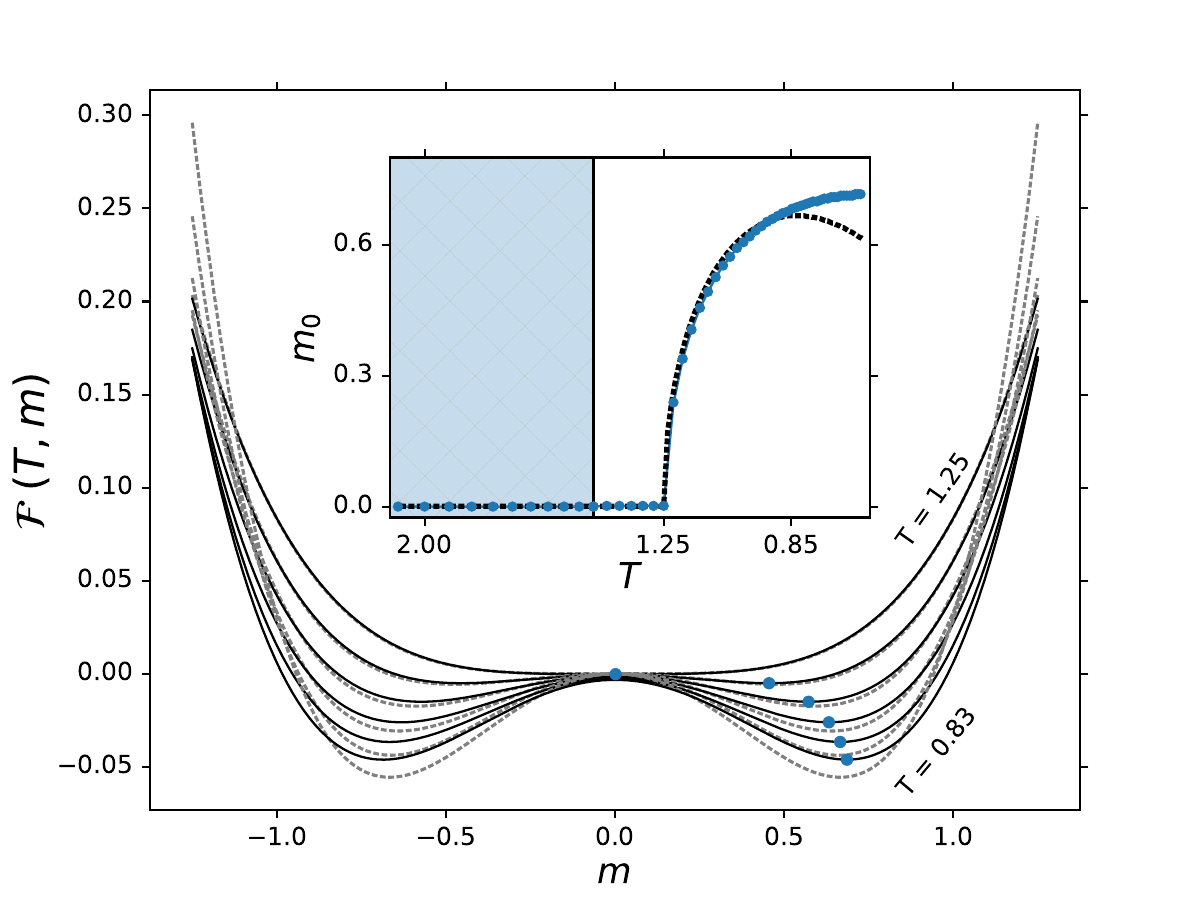}
	\caption{
	{{GP prediction (solid curves) of the free energy density $f(T,m)$ of the mean-field Heisenberg model produced by Eq.  (\ref{mf_H}) (dashed curves).
	Inset: the order parameter $m_0$ that minimizes $f(T,m)$: symbols -- GP predictions, dashed curve -- from Eq. (\ref{mf_H}). 
	The GP models are trained with 330 points at $1.47 <T < 2.08$ (shaded area) and $-1.25 <m < 1.25$. 
		}}}
	\label{heisenberg}
\end{figure}

As a third independent test, we applied the method to the Holstein polaron model defined by Eq. (\ref{polaron}) with $V_{\rm{e-ph}} = {\rm const} \sum_{k,q} c^\dagger_{k+q}c_k \left ( b^\dagger_{-q}  + b_q\right )$. Such model is known to have no transitions \cite{no-transition}. We find that the method presented here can extrapolate accurately the polaron dispersions to a wide range of the Hamiltonian parameters and yields predictions that exhibit no sign of transitions. {Since it is often not feasible to explore the entire phase diagram with rigorous quantum calculations, especially for models with many independent parameters, predicting the absence of transitions is as important as locating different phases.}

 {Finally, we demonstrate the method on an analytically soluble model. We consider the Heisenberg model $H = -\frac{J}{2}\sum_{i,j}\vec{S}_i.\vec{S}_j$ in the nearest-neighbor approximation. Employing a mean-field description, the resulting free energy density at temperature $T$ is \cite{Chaikin, sachdev, SuppT}
\begin{eqnarray}
f(T,m) \approx \frac{1}{2}\left (1- \frac{T_c}{T} \right) m^2 + \frac{1}{12}\left (\frac{T_c}{T} \right)^3 m^4,
\label{mf_H}
\end{eqnarray}
where $m$ is the magnetization and $T_c = 1.25$ the critical temperature of the phase transition. ${T} > T_c$ corresponds to the paramagnetic phase, while ${T} < T_c$ is the ferromagnetic phase.

We train GP models by the results of Eq. (\ref{mf_H}) in the paramagnetic phase far away from $T_c$ (shaded region in the inset of Figure \ref{heisenberg}).
We then extrapolate the function $f(T,m)$ across the critical temperature and compute the order parameter $m_0$ which minimizes $f(T,m)$. 
Figure \ref{heisenberg} demonstrates that $m_0$ thus predicted can be accurately extrapolated across $T_c$ and far into a different phase. 
This demonstrates again the general idea behind the technique developed here: use ML to predict the {\it evolution} of continuous functions that encodes phase transitions.

}

{It is important to point out that the iterative kernel selection algorithm of Figure 1 must be analyzed before the present method is used for the {\it quantitative} extrapolation. 
As the iterations continue, the kernels become more complex, more prone to overfitting and more difficult to optimize. 
The quantitative accuracy of the prediction may, therefore, decrease. 
The Supplemental Material illustrates the convergence to Figures 2, 3 and 4 with the kernel optimization levels and also the increase of the prediction error after a certain number of levels. 
To prevent this problem, we stop the kernel optimization when the prediction error is minimal, as explained in the Supplemental Material. 
We emphasize that this does not affect the prediction of the transitions: once a certain level of Figure 1 is reached, kernels from the subsequent optimization levels predict the transitions. We have confirmed this for all the results (Figures 2, 3 and 4) presented here. 
Thus, if the goal is to predict the presence or absence of transitions, this method can be used without validation. It is sufficient to check that subsequent levels of the kernel optimization do not produce or eliminate transitions. In order to predict quantitatively the quantum properties by extrapolation, the training data must be divided into the training and validation sets. The models must then be trained with the training set and the error calculated with the validation set. The kernel optimization must then be stopped, when the error is minimal. This is a common approach to prevent the overfitting problem in ML with artificial neural networks. 
}

{\it Summary.} We have presented a powerful method for predicting sharp transitions in Hamiltonian phase diagrams by extrapolating the properties of quantum systems. The method is based on Gaussian Process regression with a combination of kernels chosen through an iterative procedure maximizing the predicting power of the kernel. 
The model thus obtained captures the change of the quantum properties as the system approaches the transition, allowing the extrapolation of the physical properties, even across sharp transition lines.

  We believe that the present work is the first example of the application of ML for extrapolation of physical observables for quantum systems. 
  We have demonstrated that the method is capable of using the properties of the quantum system within a given phase to predict not only the closest sharp transition, but also a transition removed from the training points by a separate phase. This makes the present method particularly valuable for searching phase transitions in the parts of the parameter space that cannot be probed experimentally or theoretically. {Given that the training of the models and the predictions do not present any numerical difficulty \cite{SuppNumerics}}, the present method can also be used to guide rigorous theory or experiments in search for phase transitions. {Finally, we must note that, although the present extrapolation method works well for all four problems considered, we cannot prove that it is accurate for an arbitrary system so the predictions must always be validated, as is common in machine learning.}
  

\begin{acknowledgments}
We acknowledge useful discussions with Dries Sels and Mathias Sheurer. This
work was supported by the Natural Sciences and Engineering Research Council
of Canada (NSERC) and by a visiting student fellowship at the Institute for
Theoretical Atomic, Molecular, and Optical Physics (ITAMP) at Harvard
University and the Smithsonian Astrophysical Observatory (J.S.).
\end{acknowledgments}



\end{document}



\title{Supplemental Material for\\
``Extrapolating quantum observables with machine learning: Inferring multiple phase transitions from properties of a single phase"
}



\author{Rodrigo A. Vargas-Hern\'andez}
\address{Department of Chemistry, University of British Columbia, Vancouver, British Columbia, Canada, V6T 1Z1
}
\author{John Sous}
\address{Department of Chemistry, University of British Columbia, Vancouver, British Columbia, Canada, V6T 1Z1
}
\address{Department of Physics and Astronomy, University of British Columbia, Vancouver, British Columbia, Canada, V6T 1Z1}
\address{Stewart Blusson Quantum Matter Institute, University of British Columbia, Vancouver, British Columbia, Canada, V6T 1Z4}
\author{Mona Berciu}
\address{Department of Physics and Astronomy, University of British Columbia, Vancouver, British Columbia, Canada, V6T 1Z1}\address{Stewart Blusson Quantum Matter Institute, University of British Columbia, Vancouver, British Columbia, Canada, V6T 1Z4}
\author{Roman V. Krems}
\address{Department of Chemistry, University of British Columbia, Vancouver, British Columbia, Canada, V6T 1Z1
}


\date{\today}

\begin{abstract}

\end{abstract}

\maketitle

The purpose of this supplemental material is to provide details of the numerical calculations we present in this work. Sections I and II discuss the machine-learning methods and Sections III -- the quantum calculations used to train the ML models. 


\section{GP regression with kernel combinations}
Gaussian process (GP) regression is a kernel-based probabilistic non-parametric supervised ML algorithm \cite{gpbook}. 
Within the GP regression framework, the prediction is a normal distribution characterized by a mean $\mu(\cdot)$ and a standard deviation $\sigma(\cdot)$, given as
\begin{eqnarray}
\mu(\bm x_*) &=& K({\bm x_*},\bm x)^\top \left [ K(\bm x, \bm x) + \sigma_n^2 I \right ] ^{-1}{\bm y} \label{eqn:gp_mu}\\
\sigma(\bm x_*) &=& K({\bm x_*},{\bm x_*}) - K({\bm x_*},\bm x)^\top\left [ K(\bm x,\bm x) + \sigma_n^2 I\right ]^{-1}K({\bm x_*},\bm x).
\end{eqnarray} 
Here, 
\begin{itemize}
\item  $\bm x = (\bm x_1, \bm x_2, ..., \bm x_n)^\top$  is a vector of $n$ points in a multi-dimensional parameter space, where the GP model is trained;
\item $\bm x_i$ is a vector of variable parameters. \\ \\
For the case of the polaron models considered here, \\ {\color{white} a} \hspace{-0.1cm} $\bm x_i \Rightarrow \{ {\rm polaron~momentum~}K,~ {\rm Hamiltonian~parameter}~\alpha, ~ {\rm Hamiltonian~parameter}~\beta, {\rm phonon~frequency~}\omega\}$. \\
For the case of the Heisenberg model considered here, {$\bm x_i \Rightarrow \{ {\rm Temperature}~T,~ {\rm magnetization}~\tilde{m} \}$};
\item $\bm y = f(\bm x)$ is a vector of quantum mechanics results at the values of the parameters specified by $\bm x_i$
\\ \\
For the case of the polaron models considered here, $\bm y \Rightarrow {\rm polaron~energy~}E$. \\
For the case of the Heisenberg model considered here, $\bm y \Rightarrow {\rm free~energy~density}$; 
\item $\bm x_\ast$ is a point in the parameter space where the prediction $\bm y_\ast$ is to be made;
\item $K(\bm x, \bm x)$ is the $n \times n$ square matrix with the elements $K_{i,j} = k(\bm x_i,\bm x_j)$ representing the covariances between $\bm y(\bm x_i)$ and $\bm y(\bm x_j)$. The  elements $k(\bm x_i,\bm x_j)$ are represented by the kernel function. 
\end{itemize}

The GP models are constructed (in the language of ML ``trained'') by the quantum mechanics results $\bm y$ at the parameters in $\bm x$. The unknown in this model is the kernel function. The goal of the training is thus to find the best representation for the kernel function $k(\cdot, \cdot)$. 

In a standard procedure for training a GP model, one begins by assuming some simple analytical functional form for $k(\cdot, \cdot)$. For example, one assumes {\it one} of the following functional forms:
\begin{eqnarray}
k_{\rm LIN}(\mathbf{x}_i, \mathbf{x}_j)  &=&  \mathbf{x}_i^\top \mathbf{x}_j
\label{eqn:k_LIN}\\
k_{\rm RBF}(\mathbf{x}_i, \mathbf{x}_j) &=& \exp \left(-\frac{1}{2}r^2(\mathbf{x}_i, \mathbf{x}_j)\right)
\label{eqn:k_RBF}\\
k_{\rm MAT}(\mathbf{x}_i, \mathbf{x}_j)  &=& \left( 1 + \sqrt{5}r^2(\mathbf{x}_i,\mathbf{x}_j) +  \frac{5}{3}r^2(\mathbf{x}_i, \mathbf{x}_j)\right )\nonumber \\
&&\times \exp\left ( -\sqrt{5}r^2(\mathbf{x}_i, \mathbf{x}_j)\right )~~~~
\label{eqn:k_MAT}\\
k_{\rm RQ}(\mathbf{x}_i, \mathbf{x}_j)  &=& \left ( 1 + \frac{|\mathbf{x}_i- \mathbf{x}_j|^2}{2\alpha\ell^2} \right )^{-\alpha}
\label{eqn:k_RQ}
\end{eqnarray}
where $r^2(\mathbf{x}_i, \mathbf{x}_j) = (\mathbf{x}_i- \mathbf{x}_j)^\top \times {M} \times (\mathbf{x}_i-\mathbf{x}_j)$ and ${M}$ is a diagonal matrix with different length-scales $\ell_d$ for each dimension of $\mathbf{x}_i$. This list represents some of the most commonly used kernel functions.

The parameters of this analytical form are then found 
by maximizing the log \emph{marginal likelihood} function,
\begin{eqnarray}
\log p(\bm{y}|\bm x,\bm{\theta}) = -\frac{1}{2} \bm{y}^\top K^{-1} \bm{y} - \frac{1}{2}\log |K| -\frac{n}{2} \log (2\pi),
\label{eqn:logml}
\end{eqnarray}
where $\theta$ denotes collectively the parameters of the analytical function for $k(\cdot, \cdot)$ and $|K|$ is the determinant of the matrix $K$.

The marginal likelihood can also be used as a metric to compare different kernels. However, care must be taken when kernels with different numbers of parameters are to be compared. The second term of Eq. (\ref{eqn:logml}) directly depends on the number of parameters in the kernel, which makes the log marginal likelihood inappropriate to compare kernels with different numbers of parameters.
To overcome this issue, we compare the predictive power of different kernels using the Bayesian Information criterion (BIC) \cite{bic}
\begin{eqnarray}
\text{BIC}({\cal M}_i) = \log p(\mathbf{y} | {\bm x}, \hat{\mathbf{\theta}}, {\cal M}_i) -\frac{1}{2}|{\cal M}_i|\log n
\label{BIC-eq}
\end{eqnarray}
where $|{\cal M}_i|$ is the number of kernel parameters of the kernel ${\cal M}_i$. Here, $ p(\mathbf{y} | {\bm x}, \hat{\mathbf{\theta}}, {\cal M}_i)$ is the marginal likelihood for the optimized kernel $\hat{\mathbf{\theta}}$ which maximizes the logarithmic part. 
The last term in Eq. (\ref{BIC-eq}) acts to penalize kernels with larger number of parameters to reduce overfitting, thus making the predicting model more robust.

\subsection{Learning with kernel combinations}

Typically, GP regression is used for interpolation of the training points $\bm y(\bm x_i)$. For this problem, it is sufficient to choose one of the kernel functions above. The choice of the function will determine the efficiency of the interpolation model, i.e. the number $n$ of training points required for accurate predictions between the training points. However, any of the kernel functions will work for interpolation. 

As discussed in the main text, this is not the case for extrapolation. For accurate extrapolation, one needs to increase the complexity of kernels in order to capture the physical behaviour of the training data $\bm y(\bm x_i)$ well. However, complex kernels come with a risk of overfitting. In addition, the ambiguity as to the choice of the kernel function increases with the complexity of the kernel function. So, the question is, how to increase the complexity of the kernel functions in a systematic way that prevents overfitting and results in a model that captures the physical behaviour of the training results? 

Ideally, one should choose a kernel function that captures all of the physical behaviour of the training data. 
However, as we explained above, hand-crafting the `best' kernels is not an easy task \cite{gpbook,gp-ss}. In addition, hand-crafting kernels may introduce biases, 
limiting the generality of the prediction. In this work, we do not use any prior information for the selection of the kernels and we do not hand-craft kernels. 



Here, we follow Refs. \cite{kernel_comb,gp-ss} to increase the complexity of kernels by combining the simple functions (\ref{eqn:k_LIN}) - (\ref{eqn:k_RQ}) into products and sums. 
Combining different kernels can enhance the learning capacity of the GP regression \cite{kernel_comb}.


For example, the first kernel combination that we describe here is the addition of two kernels like $k_{MAT} + k_{MAT}$ or $k_{RQ} + k_{MAT}$. This new type of kernel form is capable of learning long-range and short-range correlations between data points.
Multiplication of kernels is also another possible algebraic operation, for example, $k_{RQ} \times k_{MAT}$. Multiplying any of the kernels by the linear kernel, {\em e.g.}, $k_{RBF}\times k_{LIN}$, leads to a GP regression that can learn increasing variations of the data.
The dot-product/linear kernel, Eq. (\ref{eqn:k_LIN}), can be used to construct polynomial kernels. For example, to describe quadratic functions, one could multiply this kernel by itself: $k_{LIN} \times k_{LIN}$.

It becomes clear that combining kernels in GP regression can provide an advantage in describing a variety of mathematical functions to accurately make predictions. This is the basis behind using GP regression with kernel combinations for extrapolating observables. To build more robust and flexible GP models, we employ the greedy search algorithm and the BIC to algorithmically construct the 'optimal' model. The greedy search is an `optimal policy' algorithm \cite{RL} that selects the kernel assumed optimal based on the BIC at every step in the search. The underlying assumption is that the BIC represents the optimal measure of the kernel performance.

{
The number of free parameters for each of the simple kernels used in this work are
\begin{itemize}
\item $k_{LIN}(\mathbf{x}_i,\mathbf{x}_j)  \Rightarrow 1$
\item $k_{RBF}(\mathbf{x}_i,\mathbf{x}_j)  \Rightarrow d$
\item $k_{MAT}(\mathbf{x}_i,\mathbf{x}_j) \Rightarrow d$
\item $k_{RQ}(\mathbf{x}_i,\mathbf{x}_j) \Rightarrow 2$,
\end{itemize}
where $d$ is the dimensionality or number of features of the data. For example, for the results presented in Figure 4 of the main text $d= 2$, $\mathbf{x}_i = [T,m]$.

Every kernel considered in this work is scaled by the constant kernel, $k_{c}(\mathbf{x}_i,\mathbf{x}_j) = const$. 
The total number of parameters of a GP model with any simple kernel considered in this work is thus increased by one due to $k_{c}(\mathbf{x}_i,\mathbf{x}_j) \times k_{X}(\mathbf{x}_i,\mathbf{x}_j)$, where $k_{X}$ is any of the kernels listed above.


As the algorithm depicted in Figure 1 progresses to lower levels, the number of free kernel parameters increases and 
 the kernels become rather complex. 
We express such kernels as the sum of products of kernels by distributing all products of sums. For example, the kernel used to construct Figure 3 (lower panel) of the main text is,
\begin{eqnarray}
(k_{MAT} \times k_{LIN} + k_{RBF}) \times k_{LIN} = k_{MAT} \times k_{LIN} \times k_{LIN} + k_{RBF} \times k_{LIN}, \nonumber
\end{eqnarray} 
which including the constant kernel is,
\begin{eqnarray}
 \big(k_{c} \times k_{MAT} \times k_{LIN} \times k_{LIN}\big) +  \big(k_{c} \times k_{RBF} \times k_{LIN}\big). \nonumber
\end{eqnarray}
}

\subsection{Numerical difficulty of training and using the GP models}

In order to train a GP model, one needs to maximize the log-likelihood function in Eq. (\ref{eqn:logml}) by iteratively computing the inverse and the determinant of the correlation matrix $K$. The dimension of this matrix is $n \times n$, where $n$ is the number of training points. In this work, $n \approx 200 - 1000$, as discussed in the next section.  Therefore, training a single GP model presents no numerical difficulty and typically takes seconds to minutes of CPU time. 

In order to find the optimal kernels using the algorithm depicted in Figure 1 of the main manuscript, one needs to train many GP models. As the levels in Figure 1 become deeper, kernels become more complex and the algorithm requires the iterative construction of more GP models. For levels 5 to 10 of Figure 1, the kernel optimization may take up to a few hours of CPU time on a single compute core. 

Using the model to predict the quantum properties involves the evaluation of the vector - matrix product in Eq. (\ref{eqn:gp_mu}). The size of the vector $n$ and the dimension of the matrix is $n \times n$, where $n \approx 200 - 1000$, as before. (Since the matrix $K$ is, at this point known, the prediction may actually be reduced to a scalar product of two vectors of size $n$). The numerical evaluation of these products presents no computational difficulty.

\section{Specific details of the extrapolation method}

The value of a quantum observable depends on the parameters of the Hamiltonian. One can learn the behavior of quantum observables using different ML models using the following relation
\begin{eqnarray}
E = \langle  \hat{{\cal H}} (K,\alpha,\beta,\dots) \rangle \sim {\cal F}(K,\alpha,\beta,\dots)
\end{eqnarray}
where ${\cal F}(\cdot)$ is any ML model that can learn the dependence between the Hamiltonian parameters and the quantum observable. In the present work, the quantum observables are the polaron ground state energy and the free-energy density denoted as $E(\cdot)$.
We use the algorithm proposed above to learn ${\cal F}(\cdot)$ and hence to extrapolate quantum observables. 

The results of Figure 2 of the main text are for the polaron model with $\beta = 0$. This figure presents the extrapolation with three different ML models, represented by circles, triangles and pentagons. 

For the predictions represented by triangles: 
\begin{itemize}
\item The GP model is  trained with 210 points distributed in the ranges $0 \leq K \leq \pi$, $0 \leq \lambda_{SSH} \leq 0.5$
\end{itemize}

For the predictions represented by circles: 
\begin{itemize}
\item The GP model is  trained with 245 points distributed in the ranges $0 \leq K \leq \pi$, $0 \leq \lambda_{SSH} \leq 0.6$
\end{itemize}

For the predictions represented by pentagons: 
\begin{itemize}
\item The GP model is  trained with 175  points distributed in the ranges $0 \leq K \leq \pi$, $0 \leq \lambda_{SSH} \leq 0.4$
\end{itemize}

The results of Figure 3 of the main text are for the polaron model with $\alpha \neq 0$ and $\beta \neq 0$. This figure presents the extrapolation with the ML models, trained by the quantum calculations at the Hamiltonian parameters shown by white circles in Figure 3 of the main text. For each training point (each white circle), we use $16$ energy points in in the range $0 \leq K \leq \pi$ for the total of 900 training points for the upper panel of Figure 3 and 960 training points for the lower panel of Figure 3.

\subsection{Effective mass and ground state momentum from extrapolated results}
Given the GP extrapolated $E(K)$, we compute $K_{GS}$ and $m^*$ as follows.
$K_{GS}$ is the value of the momentum that minimizes $E(K)$
\begin{eqnarray}
K_{GS}(\alpha, \beta, \cdots) = \argminA_K\; E(K,\alpha, \beta, \cdots)
\end{eqnarray}
which depends on the Hamiltonian parameters $\alpha$ and $\beta$.
For all results presented here, we compute $K_{GS}$ by searching for the value where $E(K)$ is minimum. This procedure is depicted in Figure SM 1.

The polaron effective mass is,
\begin{eqnarray}
m^*(\lambda_{SSH}) = \left [ \frac{\partial^2 E_{K}(\lambda_{SSH}) }{\partial K^2} \right ]^{-1} \Big|_{K=K_{GS}}
\end{eqnarray}
To compute $m^*$, we numerically evaluate the second derivative of the extrapolated $E(K)$.


\begin{figure}[h!]
        \includegraphics[width=0.5\textwidth]{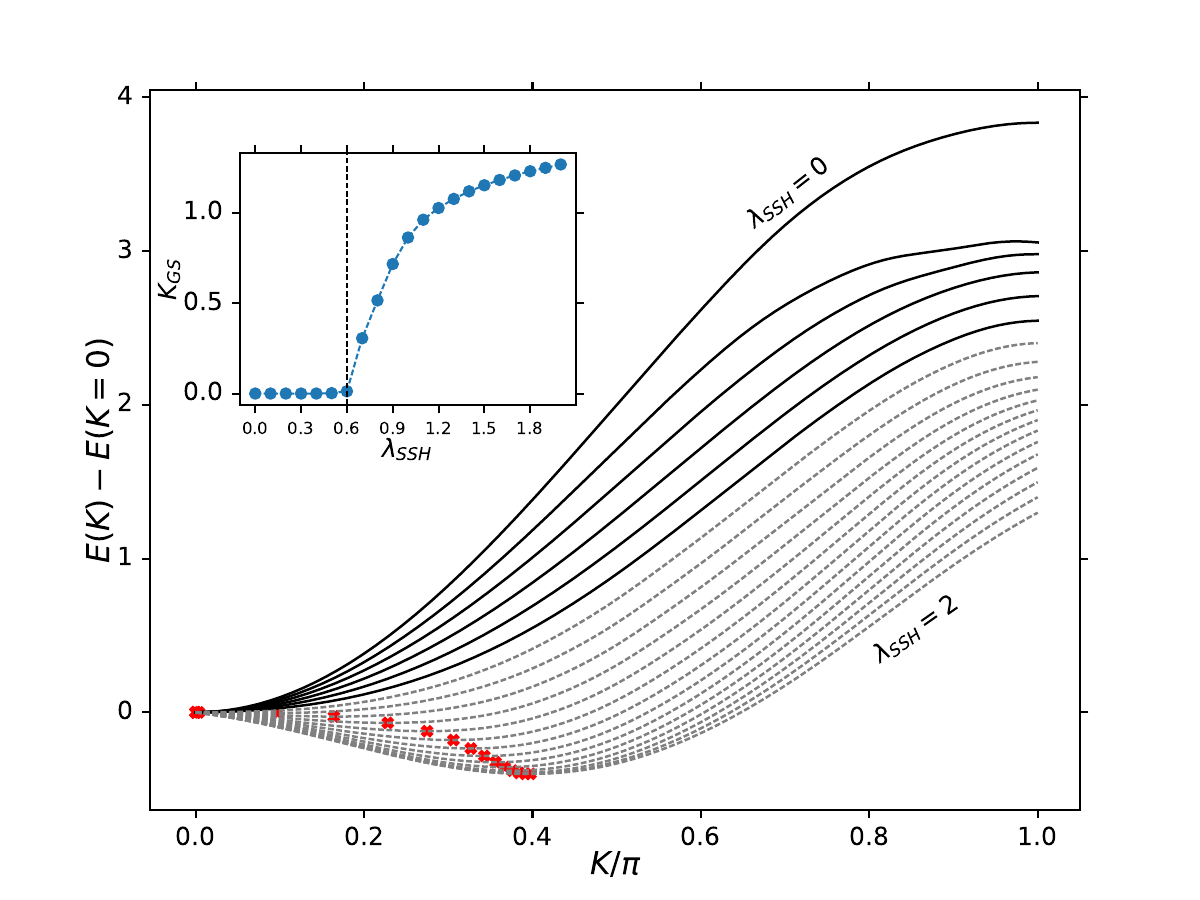}
    \caption{SSH Polaron dispersions predicted (dashed curves) with a GP model trained by the quantum calculations (black solid curves) from Ref. \cite{ssh} to result in the kernel
     $k_{RQ} \times k_{LIN} + k_{RBF}$.  The red crosses indicate the positions where the polaron dispersion reaches its minimum. Inset: the value of $K_{GS}$ as a function of $\lambda_{SSH}$.  
        \label{fig:GP-K_GS}
}
\end{figure}

\subsection{Prediction accuracy convergence (number of training points)}
 
Figure SM 2 illustrates how the accuracy of the extrapolation improves with the number of training points. We consider the pure SSH polaron model with one sharp transition. The GP models are trained by quantum results at $\lambda_{SSH}  \leq 0.4$, which is far below the transition point $\lambda_{SSH} \approx 0.6$, and used to predict the polaron properties after the transition, at $\lambda_{SSH} > 0.6$.  In all the cases presented, the kernel search algorithm depicted in Figure 1 of the main manuscript is run for three depth levels. 
 
 All models are trained by the quantum results at 5 values of $\lambda_{SSH} \leq 0.4$, but with a different number of points at $0 \leq K \leq \pi$: 15 (triangles), 25 (squares), 35 (circles). 
Figure SM 2 clearly shows that the accuracy of the prediction dramatically improves with the number of training points. 
 

\begin{figure}[h!]
        \includegraphics[width=0.5\textwidth]{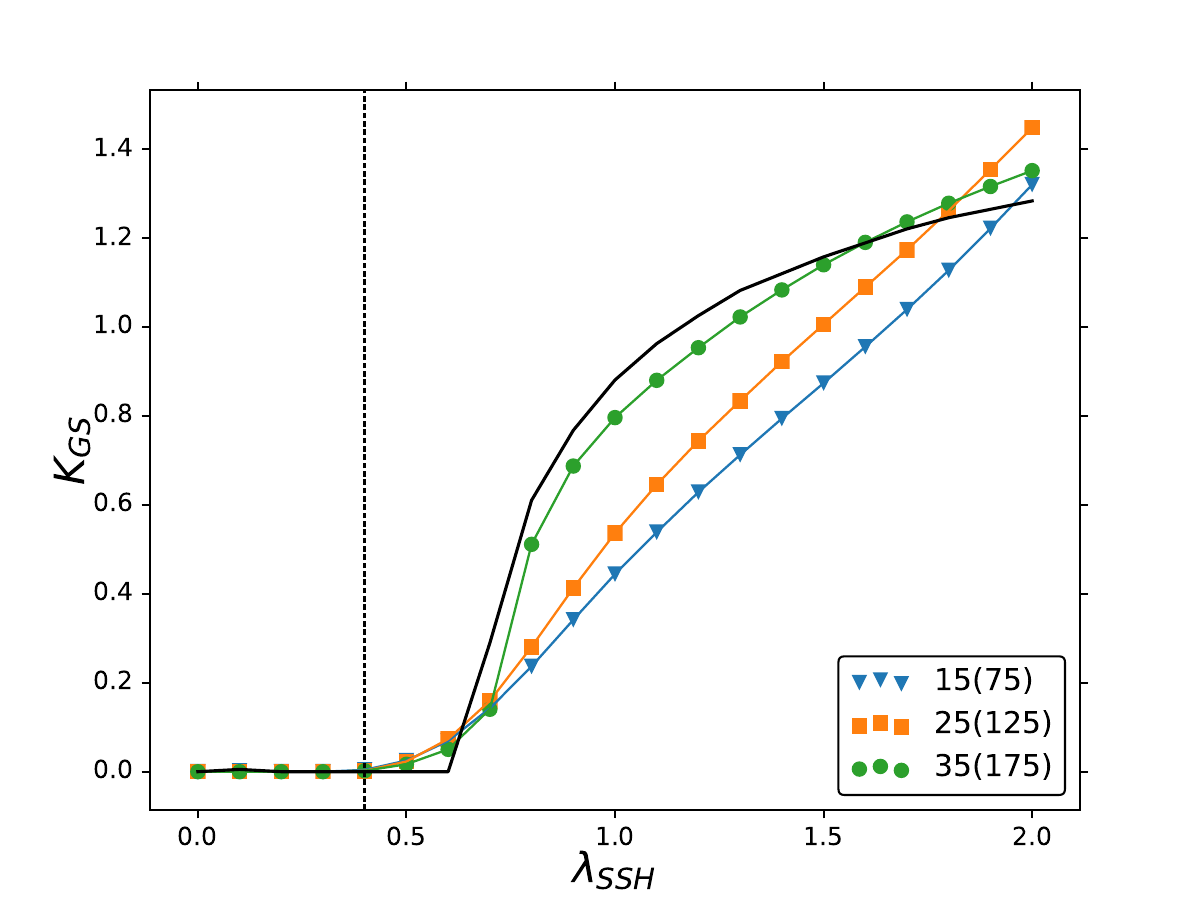}
    \caption{Ground state momentum $K_{GS}$ for the predicted SSH Polaron dispersions with a GP model trained at $\lambda_{SSH} \leq 0.4$ by thee different sets of points:  blue triangles -- 15 points per value of $\lambda_{SSH}$ (75 points total); orange squares -- 25 points per value of $\lambda_{SSH}$ (125 points total); green circles -- 35 points per value of $\lambda_{SSH}$ (175 points total).  The black solid curve is the rigorous result from Ref. \cite{ssh}. 
        \label{fig:GP-K_GS2}
}
\end{figure}

\subsection{Prediction accuracy convergence (kernel complexity dependence)}

\begin{figure}
  \includegraphics[width=.3\textwidth]{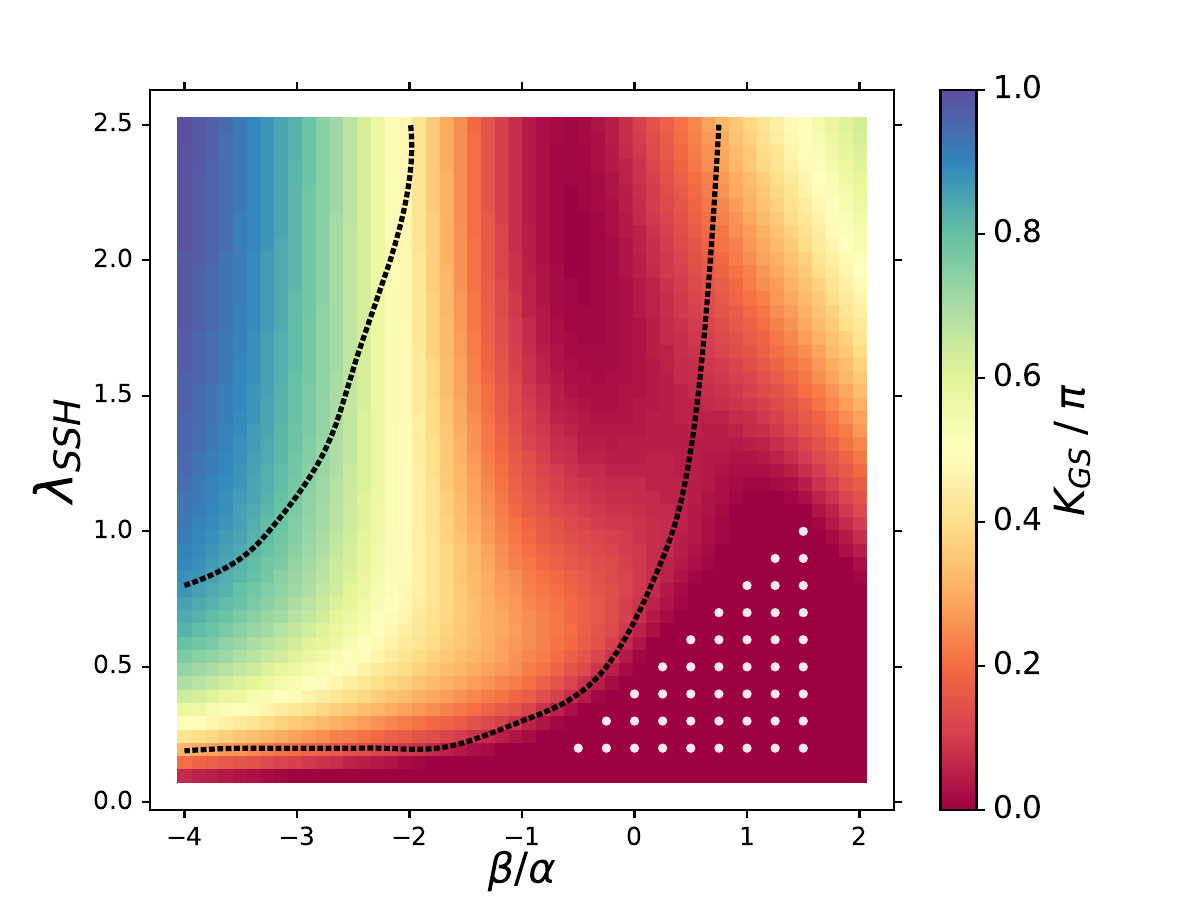}
    \includegraphics[width=.3\textwidth]{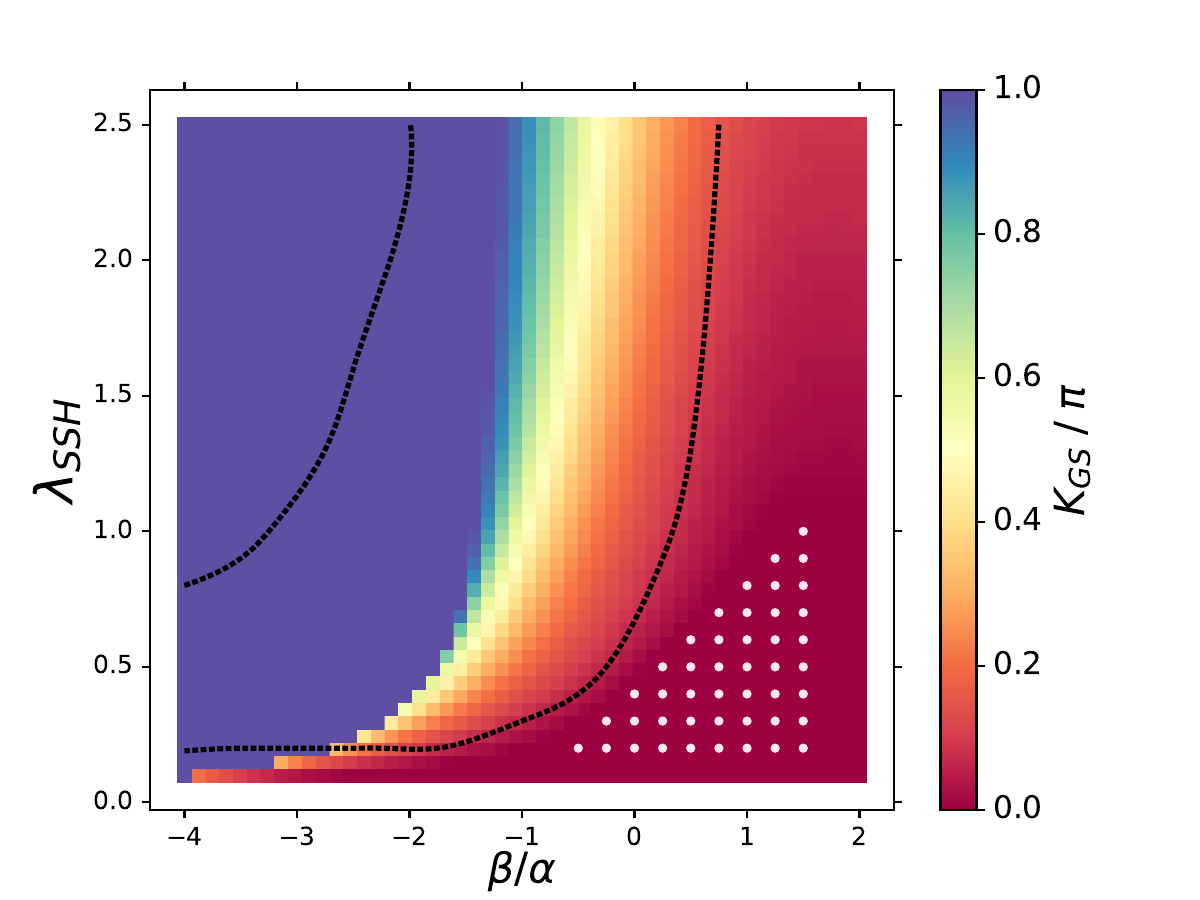}
      \includegraphics[width=.3\textwidth]{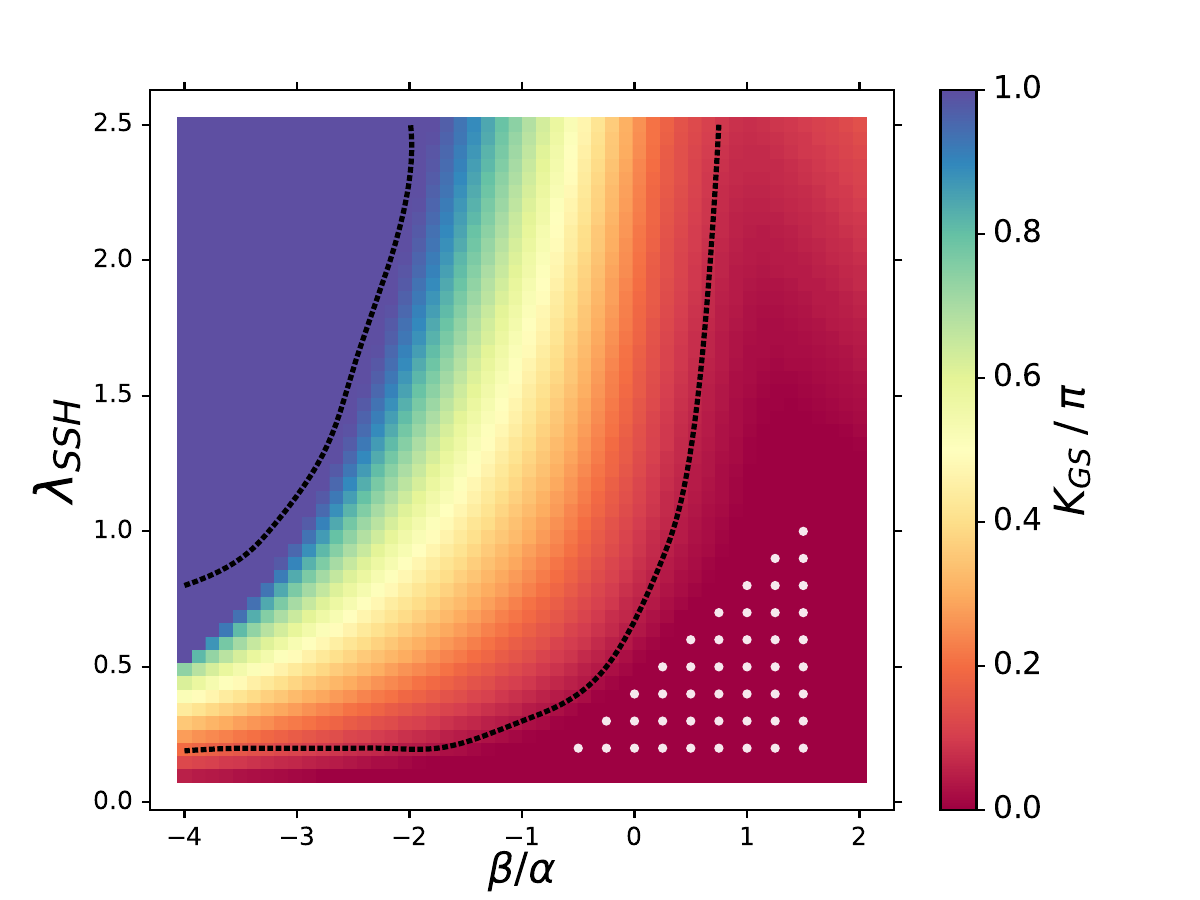}
\caption{Improvement of the phase diagram shown in Figure 3 (upper panel) of the main manuscript with the kernel complexity increasing as determined by the algorithm depicted in Figure 1 of the main manuscript. 
The panels correspond to the optimized kernels GPL-0 (left), GPL-1 (center), GPL-2 (right), where 
 ``GPL-$X$'' denotes the optimal kernel obtained after $X$ depth levels. 
   \label{fig:phase-diagram}}
\end{figure}

\begin{figure}
  \includegraphics[width=.4\textwidth]{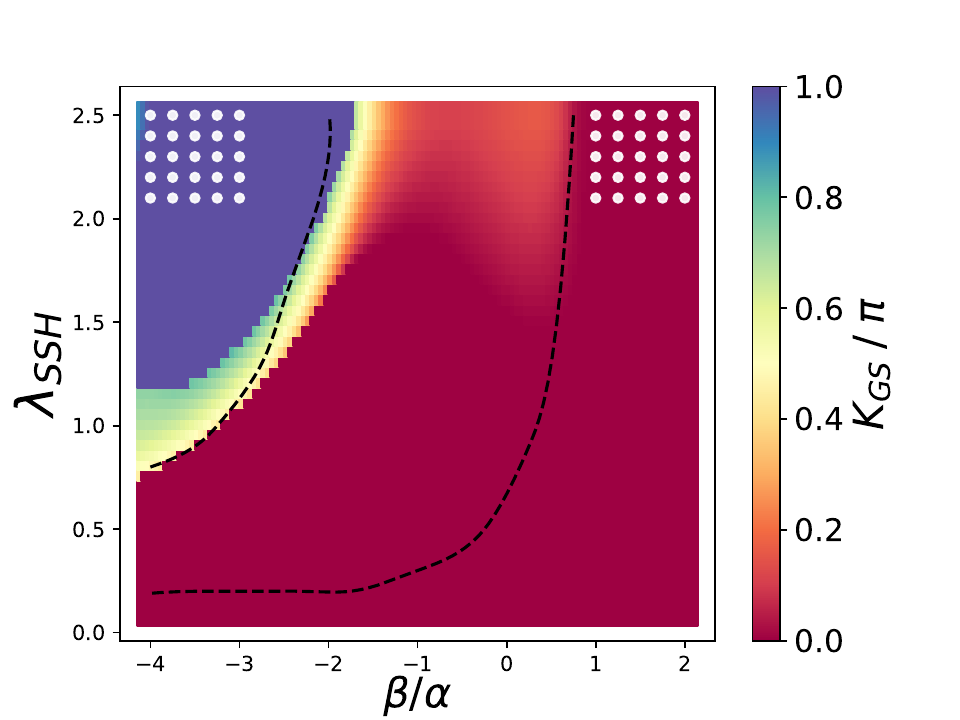}  
  \includegraphics[width=.4\textwidth]{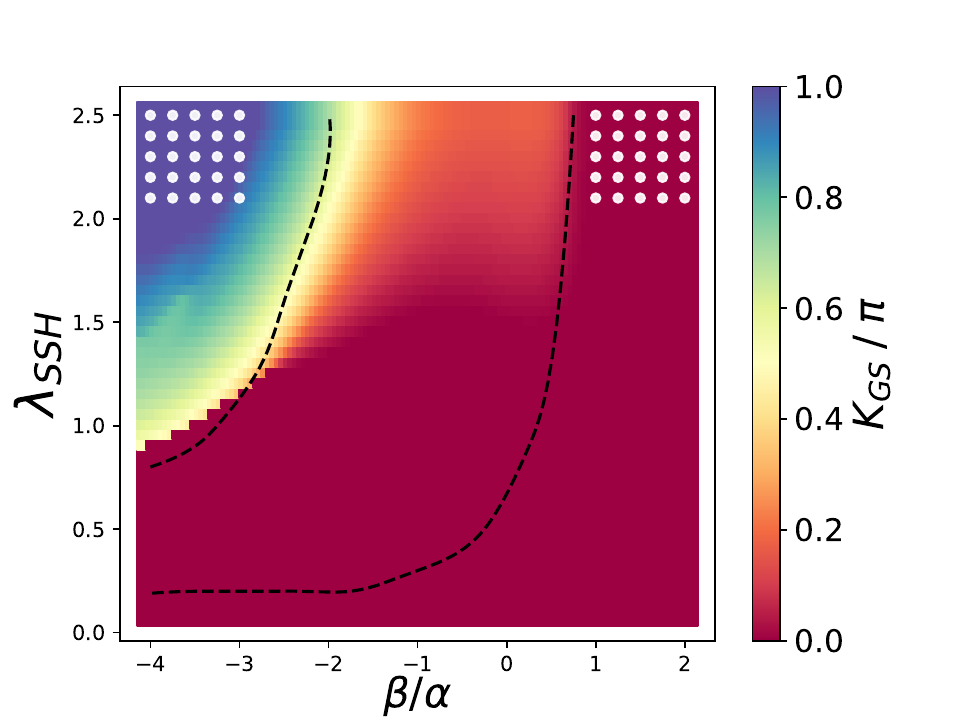}  \\
  \includegraphics[width=.4\textwidth]{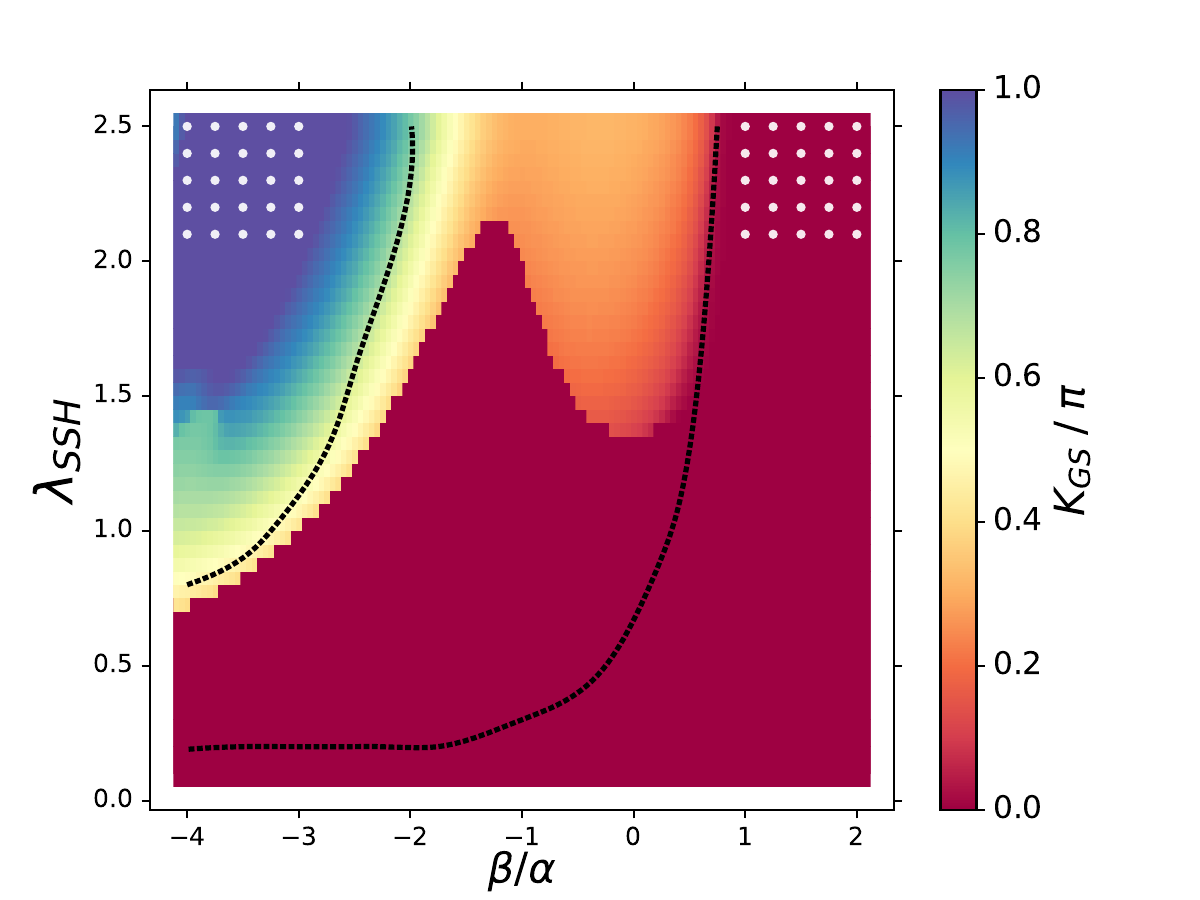}  
  \includegraphics[width=.4\textwidth]{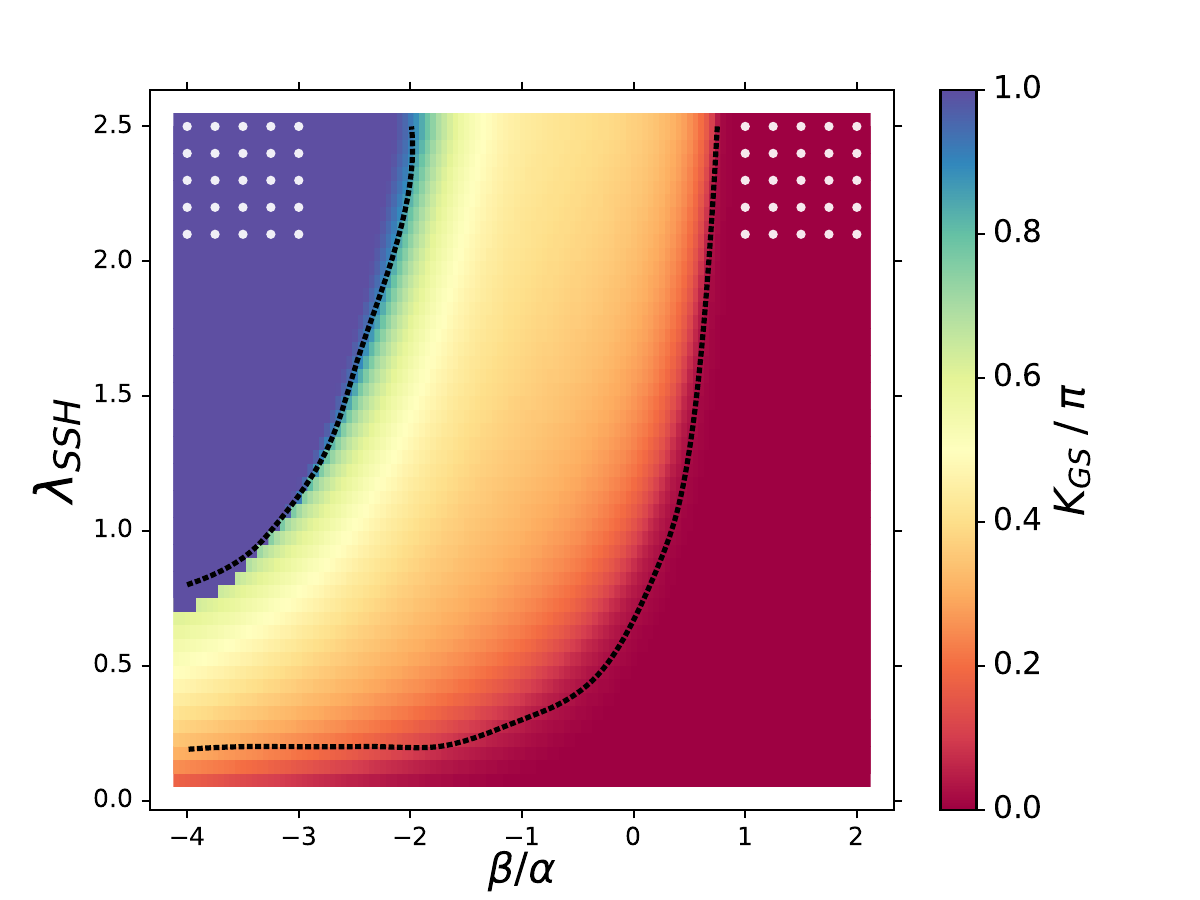}
\caption{Improvement of the phase diagram shown in Figure 3 (lower panel) of the main manuscript with the kernel complexity increasing as determined by the algorithm depicted in Figure 1 of the main manuscript. 
The panels correspond to the optimized kernels GPL-0 (upper left), GPL-1 (upper right), GPL-2 (lower left), GPL-3 (lower right), where 
 ``GPL-$X$'' denotes the optimal kernel obtained after $X$ depth levels. 
   \label{fig:phase-diagram}}
\end{figure}

For clarity, here, we use the notation ``GPL-$X$'' for the kernel with the highest BIC obtained after $X$ depth levels of the algorithm depicted in Figure 1 of the main manuscript. 
``$X$'' thus denotes the depth of the kernel optimization diagram shown in Figure 1 of the main manuscript. Figures SM 3 and SM 4 show how the accuracy of the prediction of the sharp transitions shown in Figure 3 of the main text improves as the kernel complexity increases.

For {all of the calculations presented}, we verified that increasing $X$ (the number of levels in the kernels optimization) does not change the predictions of the phase transitions. This applies to all results in Figures 2, 3 and 4 in the main manuscript as well as the Holstein model results discussed in the main text. Once a phase transition (or the absence of transitions) is identified, the prediction of the phase transition (or the absence of transitions) is reliable. In other words, once a certain level of kernel optimization is reached, all kernels from the subsequent optimization levels predict the phase transitions or the absence of the phase transitions correctly.  

 However, as the complexity of the kernels increases with each new level $X$, it becomes more difficult to find the optimal kernel within a given level $X$. The optimization algorithm is more likely to be stuck in a local minimum. This does not affect the predictions of the phase transitions. However, the quantitative predictions of the quantum properties in the extrapolated phase may be affected. Both of these points are illustrated in Figure SM 5 (upper right panel). To prevent this problem, in the present work, we stop the optimization algorithm after three levels of optimization for the results in Figures 2, 3 (upper panel) and 4. For the results in Figure 3 (lower panel), we stop the optimization after four levels.  
 
These results show that, if the goal is to predict the presence or absence of phase transitions, the method described here can be used without validation. It is sufficient to ensure that subsequent levels of the kernel optimization do not produce or eliminate phase transitions. If the goal is to predict quantitatively the quantum properties by extrapolation, the training data must be divided into a training and validation sets. The models must then be trained with the training set and the error calculated with the validation set. The kernel optimization must then be stropped at the level of the diagram in Figure 1, where the error is the smallest. This is one of the approaches to prevent the overfitting problem in machine learning with artificial neural networks. 
 

\begin{figure}
  \includegraphics[width=.4\linewidth]{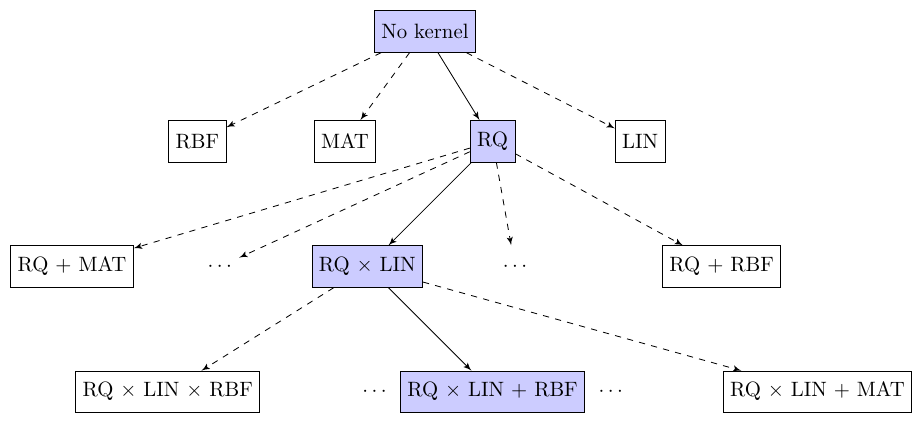}
  \includegraphics[width=0.4\linewidth]{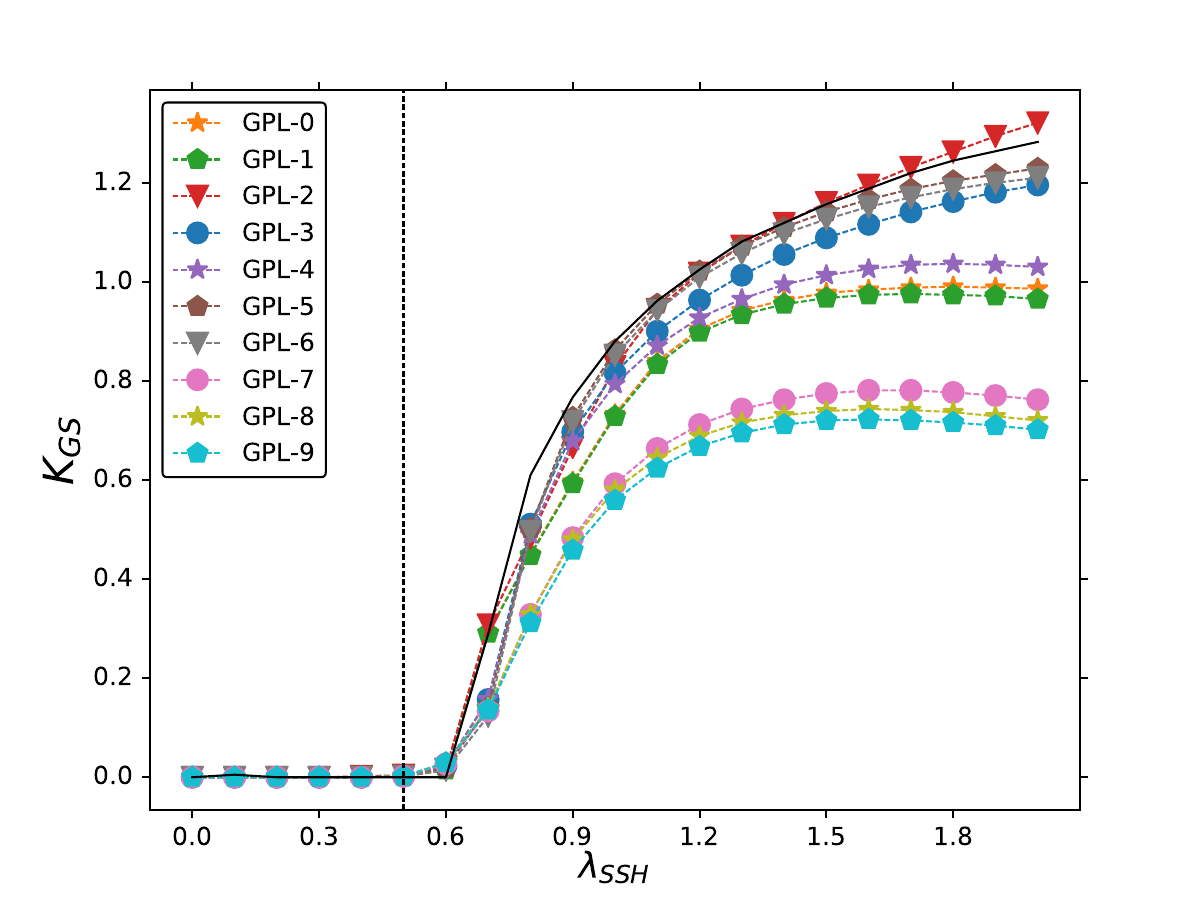}
  \includegraphics[width=0.4\linewidth]{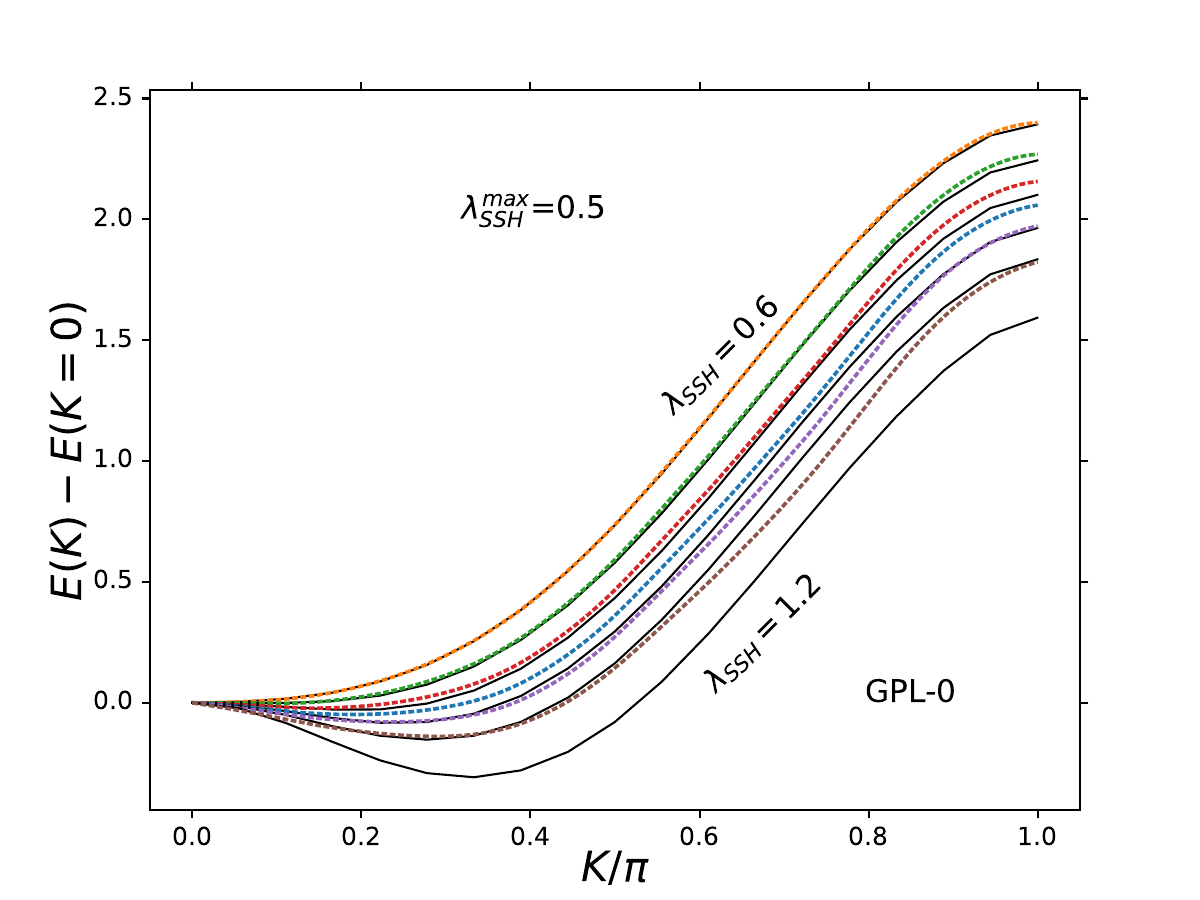}
  \includegraphics[width=0.4\linewidth]{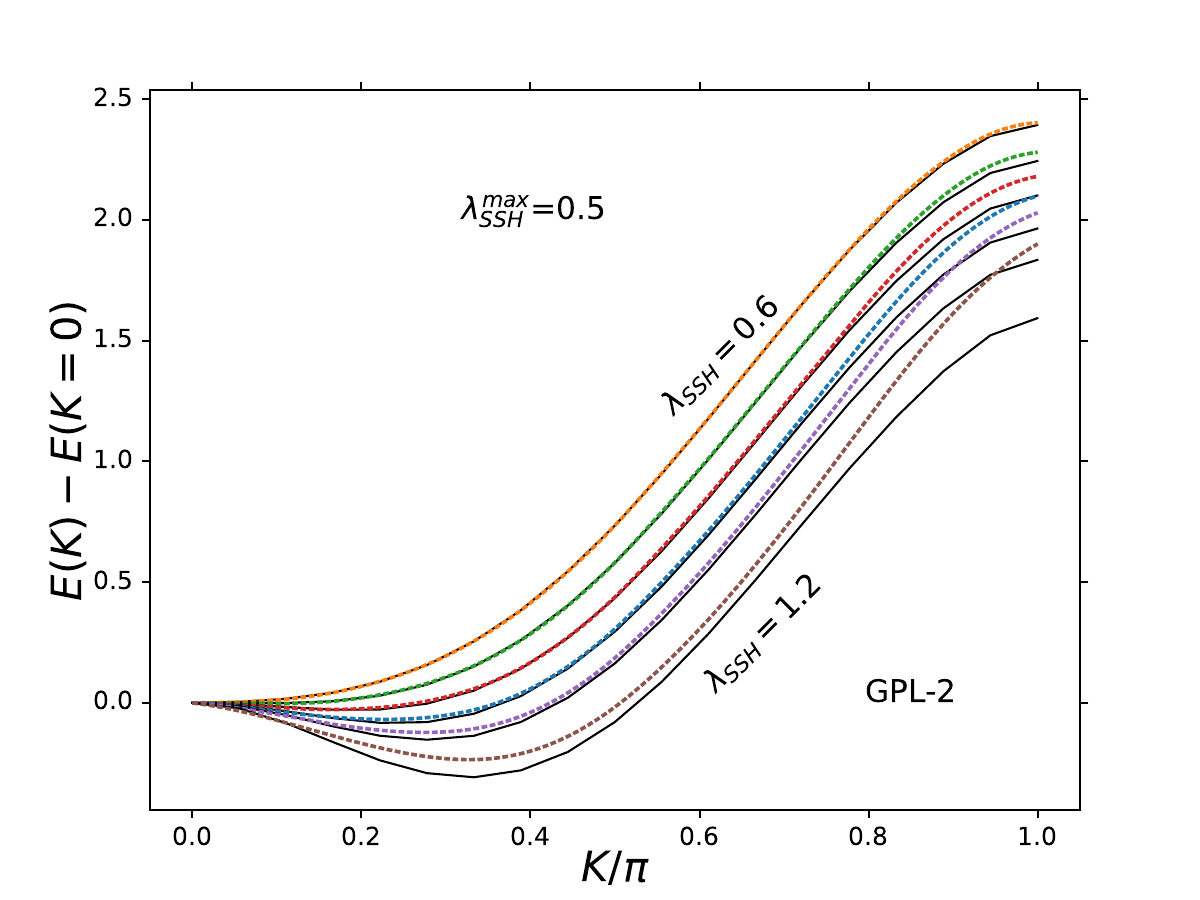}
\caption{
   Upper left: Schematic diagram of the kernel combinations with the highest BIC.
   Upper right: Effect of the increasing kernel complexity on the extrapolation accuracy. ``GPL-$X$'' denotes the results with the kernel obtained after $X$ depth levels depicted in the upper left panel (e.g. GPL-0 is $k_{RQ}$ and GPL-1 is $k_{RQ}\times k_{LIN}+k_{RBF}$). 
 Lower panels: polaron dispersions predicted by the GP model with the kernel $k_{RQ}$ (left panel) and kernel $k_{RQ}\times k_{LIN}+k_{RBF}$ obtained at the GPL-2 level (right panel). The dashed curves show the GP model predictions, while the solid curves are the results from Ref. \cite{ssh}. The GP models are trained by the quantum results at $\lambda_{SSH} \leq 0.5$.  
   \label{fig:GP-BIC}}
\end{figure}

\section{Quantum calculations to obtain training data}

\subsection{Polaron models}

For the polaron models, we use the Momentum Average (MA) approximation yielding accurate results for the polaron energies in one-dimensional lattices of infinite size \cite{BercuiPRL,BercuiPRB,GoodvinPRB}. 

The MA approach is a non-perturbative quasi-analytical technique designed to solve the equation of motion for the Green's function $\hat{G}(k,\omega) = \bra{k}(\omega - \hat{H} + i\eta)^{-1}\ket{k}$. We use the Dyson's identity $\hat{G}(\omega) = \hat{G}_0(\omega) +\hat{G}(\omega)\hat{V}\hat{G}_0(\omega)$ to generate the hierarchy of equations of motion. $\hat{G}(\omega) = (\omega - \hat{H} + i\eta)^{-1}$, $\hat{G}_0(\omega) = (\omega - \hat{H}_0 + i\eta)^{-1}$ with $\hat{H}_0 = \hat{H} - \hat{V}_{\rm{e-ph}} $, and $\hat{V} = \hat{V}_{\rm{e-ph}}$ is the electron-phonon coupling term. This hierarchy consists of an infinite set of coupled equations making an exact solution impossible. 

The MA approach acts to guide an insightful approximation/truncation of the hierarchy allowing for efficient yet accurate computations by neglecting the exponentially small diagrams in the expansion. The set of diagrams retained in the hierarchy is identified by considering the variational meaning of MA: one allows for boson excitations only within a finite spatial cut-off from the electron in the polaron cloud \cite{BercuiPRB}. 

This choice of the variational space depends on the details of the Hamiltonian and states of interest \cite{BercuiPRB}. For the Holstein model, a one-site phonon cloud suffices to provides accurate results for single polarons \cite{BercuiPRL,BercuiPRB} and for tightly bound bipolarons \cite{AdolphsPRB}. For the SSH model, the coupling to phonons is non-local and therefore a bigger cloud is required to yield accurate results. A three-site phonon cloud MA has been shown to be very accurate for such models \cite{ssh,FehskePRB,SousScRep,SousPRA,SousarXiv}.

By design, the MA approach computes the proprieties of polarons in infinite lattices by utilizing the momentum space representation. Therefore, finite size effects have no relevance.

The MA data used in this work are of the three-site variational flavor and have been confirmed to be in quantitative agreement with numerically exact results. The SSH polaron results were verified against the Bold Diagrammatic Quantum Monte Carlo results in Ref. \cite{ssh}, whereas more complicated extensions have been verified against the Variational Exact Diagonalization in Refs. \cite{SousScRep,SousPRA,SousarXiv}.

The polaron energy is obtained from the lowest discrete peak in the imaginary part of the Green's function.

\subsection{Mean-free energy of the Heisenberg model}
Here we present the derivation of the dimensionless mean-field free energy density of the Heisenberg model we study with the GP method to predict the transition from ferromagnetic to paramagnetic phase. The Heisenberg model Hamiltonian reads
\begin{eqnarray}
{\cal H} = -\frac{J}{2} \sum_{\langle i,j \rangle } \bar{S}_i \cdot \bar{S}_j ,
\end{eqnarray}
where $\langle i,j \rangle$ only account for nearest-neighbour interactions between different spins $\bar{S}_i$.
The free energy of the Heisenberg model in the mean-field approximation is a function of the magnetization $m$ and the temperature $T$ \cite{Chaikin},
\begin{eqnarray}
F(m,T) = \frac{JzNm^2}{2(g\mu_B)^2} - NT\ln\left [2 \cosh\left( \frac{Jzm}{2Tg\mu_B}\right ) \right ],
\end{eqnarray}
where $m$ is defined as $m = g\mu_B \langle \bar{S}_i \rangle$ and $z$ is the coordination number. The Boltzmann constant is set to 1 throughout this section. 

Taylor expanding $F(m,T)$ near the transition, where $m$ is vanishingly small, we obtain
\begin{eqnarray}
F(m,T) =  \frac{JzNm^2}{2(g\mu_B)^2}  - NT\left [ \ln(2) +  \frac{1}{2} \left(\frac{Jz}{2Tg\mu_B} \right )^2  m^2 -  \frac{1}{12} \left(\frac{Jz}{2Tg\mu_B} \right )^4 m^4 + \cdots\right ].
\end{eqnarray}
To find the critical transition temperature $T_c$, we minimize $F(m,T)$: $\frac{\partial{F}}{\partial{m}}=0$. Solving graphically, we obtain $T_c = \frac{Jz}{4}$ \cite{Chaikin}.
We then divide $F(m,T)$ by $N$ and $T_c$ after subtracting $F(0,T)$ to obtain the shifted free energy density
\begin{eqnarray}
f(m,T) = \frac{Jz}{2\left(g\mu_B \right)^2} \left [1  - \frac{T_c}{T} \right ] m^2 + \frac{4}{3\left(g\mu_B \right)^4}\left ( \frac{T_c}{T} \right )^3m^4. 
\end{eqnarray}
The last step is to define the dimensionless magnetization $\tilde{m} = \frac{2m}{g\mu_B}$, yielding
\begin{eqnarray}
f(\tilde{m},T) = \frac{1}{2} \left [1  - \frac{T_c}{T} \right ] \tilde{m}^2 + \frac{1}{12}\left ( \frac{T_c}{T} \right )^3 \tilde{m}^4 
\label{eqn:f}
\end{eqnarray}
This is Eq. (12) in the main text, where the tilde over $m$ has been omitted to simplify the notation. 

The shape of the magnetization dependence of the free energy density changes with $T$. At $ \frac{T_c}{T} < 1$, the minimizer of $f(\tilde{m},T)$, {\em i.e.} the order parameter (here denoted as $m_0$) is $m_0= 0$;  while for $ \frac{T_c}{T} > 1$, $m_0 \neq 0$. This is illustrated in Figure 4 of the main text for $Jz=5$, {\em i.e.} $T_c = 1.25$. This form of $f(\tilde{m},T)$ can be equivalently obtained through the phenomenological Landau theory of phase transitions \cite{Chaikin}. We use GP regression with kernel combinations to extrapolate the free energy density acrtoss the phase transition.
\bibliography{reference}